\documentclass[fleqn,usenatbib]{mnras}

\usepackage{newtxtext,newtxmath}

\usepackage[T1]{fontenc}
\usepackage{ulem}

\DeclareRobustCommand{\VAN}[3]{#2}
\let\VANthebibliography\thebibliography
\def\thebibliography{\DeclareRobustCommand{\VAN}[3]{##3}\VANthebibliography}

\usepackage{graphicx}	
\usepackage{amsmath}	

\title[Jet-driven outflow in IRAS 10565+2448? ]{
Does a radio jet drive the massive multi-phase outflow in the ultra-luminous infrared galaxy IRAS 10565+2448? }

\author[]{
Renzhi Su$^{1,2,3}$\thanks{E-mail: surz@shao.ac.cn},
Elizabeth K. Mahony$^{3}$,
Minfeng Gu$^{1}$,
Elaine M. Sadler$^{3,4,5}$,
S. J. Curran$^{6}$,
\newauthor
James R. Allison$^{5,7}$,
Hyein Yoon$^{4,5}$,
J. N. H. S. Aditya$^{4,5}$,
Yogesh Chandola$^{8}$,
Yongjun Chen$^{11,12}$
\newauthor
Vanessa A. Moss$^{3}$,
Zhongzu Wu$^{9}$,
Xi Shao$^{1,2}$,
Xiang Liu$^{10}$,
Marcin Glowacki$^{13}$,
Matthew T. Whiting$^{3}$,
\newauthor
Simon Weng$^{4,5}$
\\
$^{1}$Key Laboratory for Research in Galaxies and Cosmology, Shanghai Astronomical Observatory, Chinese Academy of Sciences, \\80 Nandan 
Road, Shanghai 200030, China\\
$^{2}$University of Chinese Academy of Sciences, 19A Yuquan Road, Beijing 100049, China \\
$^{3}$ATNF, CSIRO Space and Astronomy, PO Box 76, Epping, NSW 1710, Australia \\
$^{4}$Sydney Institute for Astronomy, School of Physics A28, University of Sydney, NSW 2006, Australia \\
$^{5}$ARC Centre of Excellence for All Sky Astrophysics in 3 Dimensions (ASTRO 3D) \\
$^{6}$School of Chemical and Physical Sciences, Victoria University of Wellington, PO Box 600, Wellington 6140, New Zealand \\
$^{7}$First Light Fusion Ltd., Unit 9/10 Oxford Industrial Park, Mead Road, Yarnton, Kidlington OX5 1QU, UK\\
$^{8}$Purple Mountain Observatory, Chinese Academy of Sciences, No.10 Yuanhua Road, Qixia District, Nanjing 210023, China\\
$^{9}$College of Physics, Guizhou University, 550025 Guiyang, PR China\\
$^{10}$Xinjiang Astronomical Observatory, Chinese Academy of Sciences, 150 Science 1-Street, Urumqi 830011, China\\
$^{11}$Shanghai Astronomical Observatory, Chinese Academy of Sciences, 80 Nandan Road, Shanghai 200030, China\\
$^{12}$Key Laboratory for Radio Astronomy, Chinese Academy of Sciences, Nanjing 210008, China\\
$^{13}$International Centre for Radio Astronomy Research, Curtin University, Bentley, WA 6102, Australia\\
}
\date{Accepted XXX. Received YYY; in original form ZZZ}

\pubyear{2022}

\begin{document}
\label{firstpage}
\pagerange{\pageref{firstpage}--\pageref{lastpage}}
\maketitle

\begin{abstract}
We present new upgraded Giant Metrewave Radio Telescope (uGMRT) \mbox{H\,{\sc i}} 21-cm observations of the ultra-luminous infrared galaxy IRAS 10565+2448, previously reported to show blueshifted, broad, and shallow \mbox{H\,{\sc i}} absorption indicating an outflow. Our higher spatial resolution observations have localised this blueshifted outflow, which is $\sim$ 1.36 kpc southwest of the radio centre and has a blueshifted velocity of $\sim 148\,\rm km\,s^{-1}$ and a full width at half maximum (FWHM) of $\sim 581\,\rm km\,s^{-1}$. The spatial extent and kinematic properties of the \mbox{H\,{\sc i}} outflow are consistent with the previously detected cold molecular outflows in IRAS 10565+2448, suggesting that they likely have the same driving mechanism and are tracing the same outflow. By combining the multi-phase gas observations, we estimate a total outflowing mass rate of at least $140\, \rm M_\odot \,yr^{-1}$ and a total energy loss rate of at least $8.9\times10^{42}\,\rm erg\,s^{-1}$, where the contribution from the ionised outflow is negligible, emphasising the importance of including both cold neutral and molecular gas when quantifying the impact of outflows. We present evidence of the presence of a radio jet and argue that this may play a role in driving the observed outflows. The modest radio luminosity $L_{\rm1.4GHz}$ $\sim1.3\times10^{23}\,{\rm W\,Hz^{-1}}$ of the jet in IRAS 10565+2448 implies that the jet contribution to driving outflows should not be ignored in low radio luminosity AGN.              
\end{abstract}




\section{Introduction}
Neutral atomic hydrogen, \mbox{H\,{\sc i}}, plays a significant role in galaxy formation and evolution \citep[e.g.][]{hopkins2006,dirver2018,morganti2013}. With the hyperfine transition structure, \mbox{H\,{\sc i}} can be traced either in emission or in absorption against a background radio continuum source. \mbox{H\,{\sc i}} emission has been detected in thousands of nearby galaxies \citep[e.g.][]{lang2003,meyer2004,winkel2016,catinella2018}, while \mbox{H\,{\sc i}} absorption has only been detected towards about 200 bright radio sources, see \cite{morganti2018}. One advantage of observing \mbox{H\,{\sc i}} in absorption is that the strength of the \mbox{H\,{\sc i}} absorption line has no dependence on redshift and hence can trace the \mbox{H\,{\sc i}} in the distant Universe where \mbox{H\,{\sc i}} emission is too weak to be detected with existing telescopes. Furthermore, \mbox{H\,{\sc i}} absorption experiments can be conducted with Very Long Baseline Interferometry (VLBI), making it a powerful tool in studying AGN accretion and feedback \citep[e.g.][]{maccagni2014,morganti2013,schulz2021}.   

Outflows are a key element in galaxy evolution, especially the AGN-driven outflows which can drive the co-evolution between supermassive black holes (SMBH) and their host galaxies \citep[e.g.][]{ferrarese2000,dimatteo2005,bower2006,croton2006,kormendy2013}. The interstellar medium (ISM) in galaxies is multi-phase, including ionised, neutral, and molecular gas. Previous studies of multi-phase outflows have greatly advanced our understanding of the impact of outflows \citep[e.g.][]{veilleux2005,fabian2012,cicone2014,harrison2014,mahony2013,mahony2016,heckman2017,veilleux2020,speranza2021}.

\mbox{H\,{\sc i}} absorption can reveal the neutral counterpart of outflows. A famous example is 4C\,12.50 in which a broad and blueshifted \mbox{H\,{\sc i}} outflow was found \citep{morganti2005}. Later VLBI observations spatially resolved the \mbox{H\,{\sc i}} absorption and showed that the outflow is driven by a jet \citep{morganti2013}. To date, due to the high sensitivity required, only a handful of \mbox{H\,{\sc i}} outflows have been found in bright radio sources\citep[e.g.][]{morganti2005,mahony2013,aditya2018,aditya2019}, see also a compilation by \cite{morganti2018}.

\begin{figure*}
	\includegraphics[width=15cm]{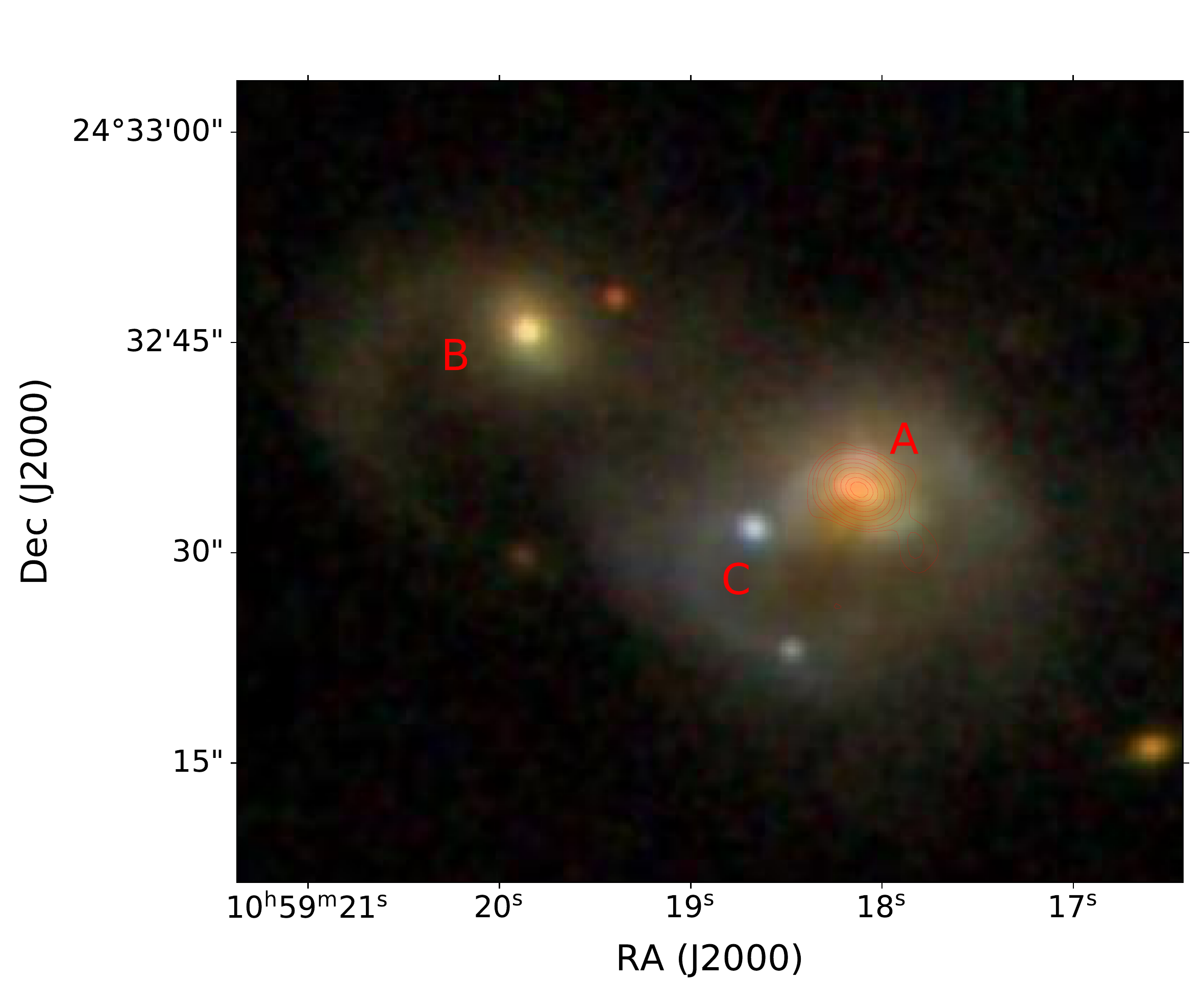}
    \caption{The SDSS $gri$ band composite image of IRAS 10565+2448. This is a merging system of three galaxies, which are marked with A, B, and C, respectively. The red contours are from the uniform-weighting radio continuum image.}
    \label{fig:opt_image}
\end{figure*}

IRAS 10565+2448 is a $z$=0.0431 ultra-luminous infrared galaxy (ULIRG) with an infrared luminosity of $1.2\times10^{12} \rm L_{\odot}$ and a star formation rate of $131.8\,\rm M_\odot$\,yr$^{-1}$ \citep{rupke2013,u2012}. It is a merging system involving three galaxies with highly disturbed stellar morphology; see the $gri$ bands composite image from Sloan Digital Sky Survey (SDSS) in Fig. \ref{fig:opt_image} on which the three galaxies have been marked as A, B, and C. 

An Arecibo observation with a velocity resolution of 8.3 km\,s$^{-1}$ and a spatial resolution of 3.3 arcmin revealed both \mbox{H\,{\sc i}} emission and absorption towards the source \citep{mirabel1988}. An intriguing \mbox{H\,{\sc i}} spectral property of IRAS 10565+2448 is a broad, shallow, and blueshifted \mbox{H\,{\sc i}} absorption wing indicating an outflow. However, due to the large Arecibo beam the \mbox{H\,{\sc i}} absorption profile is diluted by \mbox{H\,{\sc i}} emission, which precludes obtaining a clean \mbox{H\,{\sc i}} absorption profile needed to estimate the impact of the outflow. In order to disentangle the \mbox{H\,{\sc i}} emission and absorption in this system, we carried out new observations with the upgraded Giant Metrewave Radio Telescope (uGMRT). The much higher spatial resolution of this interferometer should help us obtain a clean \mbox{H\,{\sc i}} absorption line and possibly localise the \mbox{H\,{\sc i}} outflow. The presence of outflows in IRAS 10565+2448 is also supported by studies of other bands \citep[e.g.][]{rupke2005,rupke2013,cicone2014,fluetsch2019}. Throughout this paper, we use the $\Lambda$CDM cosmology with $H_{0}=70\,$km$\,$s$^{-1}\,$Mpc$^{-1}$,  $\Omega_{m}$=0.3, and $\Omega_{\Lambda}$=0.7.     

\section{Observations and data reduction}
The uGMRT observations were carried out on 31 August 2021 with a bandwidth of 50 MHz, divided into 4096 channels and centred on 1360 MHz. This setting gives a velocity coverage of -6000 $\sim$ 5000 km\,s$^{-1}$ with respect to the systemic velocity \citep{fluetsch2019} of IRAS 10565+2448 and a velocity resolution of $\sim$ 2.7 km\,s$^{-1}$. Each scan of IRAS 10565+2448 had a length of about 45 minutes and was bracketed by 5-minute scans of the phase calibrator 1111+199. The total on-source time was 6.75 hours. 3C 147 and 3C 286 were observed at the beginning, middle, and end of the entire experiment for bandpass and amplitude calibration.   

The uGMRT data was calibrated using the Astronomical Image Processing System \citep[AIPS;][]{greisen2003}. We first inspected the data set and removed any bad data arising from non-working antennas, bad baselines, time-based issues, or radio frequency interference (RFI). We then calibrated the gains and bandpass and applied the solutions. After repeating the flagging and calibrating processes for a few rounds, we masked the channels occupied by the \mbox{H\,{\sc i}} line and binned the rest for continuum imaging. We imaged the radio continuum using Difmap \citep{shepherd1997}. We cleaned the dirty image gradually. After a few rounds of phase-only and amplitude-only self calibrations, we did both amplitude and phase self calibrations with decreasing time intervals to 0.5 minutes for the solutions. We then applied the solutions obtained from self calibration to the spectral line data, which was later binned to a velocity resolution $\sim$ 21.5 km\,$\rm s^{-1}$ (every eight channels are binned). Finally, we subtracted the continuum with a linear fit applied to line-free channels and generated the spectral cube. Two different spectral-line cubes were made, with uniform and natural weighting. They have an rms noise of 0.70 mJy\,beam$^{-1}$\,channel$^{-1}$ and 0.41 mJy\,beam$^{-1}$\,channel$^{-1}$, and a resolution of 2.23 arcsec $\times$ 1.80 arcsec and 2.84 arcsec $\times$ 2.35, respectively. 

\section{Results}
\subsection{Continuum image}
\begin{figure}
	\includegraphics[width=\columnwidth]{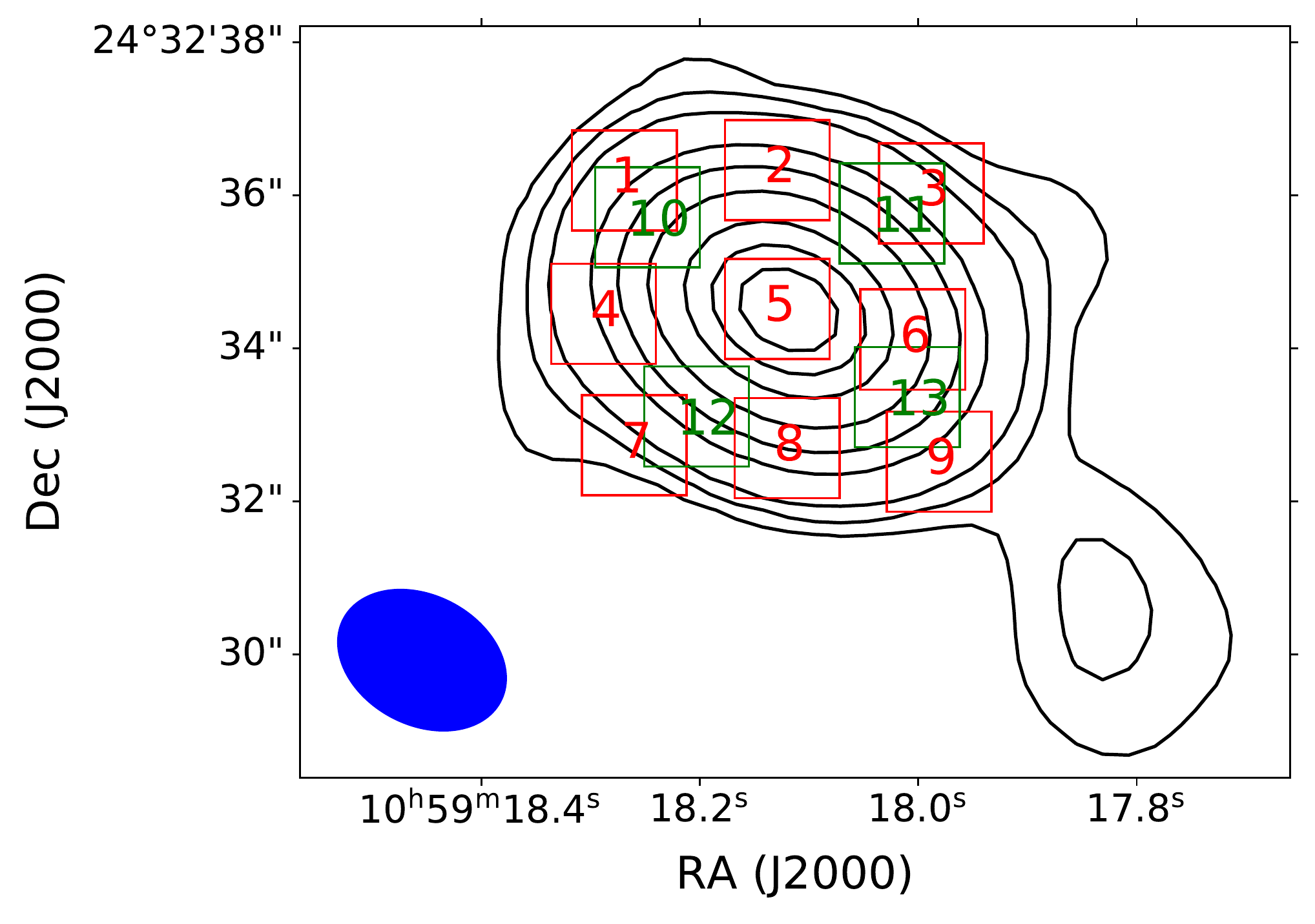}
    \caption{A radio image of galaxy A from our uGMRT observations, made with uniform weighting. The image has a resolution of 2.25 arcsec $\times$ 1.68 arcsec and a rms continuum noise of $\sim$ 27 $\mu$Jy\,beam$^{-1}$. The contours are at 0.4, 0.8, 1.5, 5, 10, 20, 40, 60, 80 percent of the peak flux density, and the beam is plotted at the lower-left corner. For the uniform-weighting and natural-weighting cubes, we distributed nine red and four green squares on the radio emission to extract spectra, respectively. Each square has a size of 1.31 arcsec $\times$ 1.31 arcsec.}
    \label{fig:radio_image}
\end{figure}

Using uniform weighting, the continuum image of IRAS 10565+2448 we obtained has a resolution of 2.25 arcsec $\times$ 1.68 arcsec and an rms noise $\sim$ 27 $\mu$Jy\,beam$^{-1}$. Galaxy B is not detected at this noise level in our observation, and the radio image of our target (galaxy A) is presented in Fig. \ref{fig:radio_image}. The radio image is resolved with a peak flux density of 39.3 mJy\,beam$^{-1}$ and a total flux density of 49.6 mJy.

\subsection{\mbox{H\,{\sc i}} lines}

\subsubsection{\mbox{H\,{\sc i}} emission}\label{emission}

Previous Arecibo observations detected both \mbox{H\,{\sc i}} emission and absorption towards IRAS 10565+2448 \citep{mirabel1988}. However, due to the low spatial resolution provided by a single dish, the \mbox{H\,{\sc i}} emission is mixed with \mbox{H\,{\sc i}} absorption and galaxies A, B and C were not able to be spatially separated. In our uGMRT observations, those galaxies were well spatially separated. Unlike the Arecibo observation, we did not detect the \mbox{H\,{\sc i}} emission in galaxy A, B or C. This is not surprising, as our cubes have a noise level similar to the faint \mbox{H\,{\sc i}} emission detected in the Arecibo observation and the uGMRT is not sensitive to the diffuse \mbox{H\,{\sc i}} emission due to the missing short spacings. In an integrated intensity map spanning velocities from -500 km\,s$^{-1}$ to -250 km\,s$^{-1}$ in the uniform-weighting cube, and a similar map spanning velocities from 250 km\,s$^{-1}$ to 500 km\,s$^{-1}$ in the natural-weighting cube, we found some pixels showing positive emission at the  $\sim$2$\sigma$ value, see the middle and right plots in Fig. \ref{fig:gas_distri}. However, by visually inspecting the mean wide-velocity spectra spanning $\sim$ -800 -- 1000 km\,s$^{-1}$, extracted from the square 14 and 17 surrounding those pixels, we concluded that these positive values arose from noise or from \mbox{H\,{\sc i}} emission that was too faint to be verified in our observation. The mean spectrum from square 17 is presented in Fig. \ref{fig:faint_HI_emi}. We suggest that more sensitive observations may help to identify the \mbox{H\,{\sc i}} emission shown on Arecibo spectrum \citep{mirabel1988}.

\subsubsection{\mbox{H\,{\sc i}} absorption}\label{absorption}

\begin{figure*}
	\includegraphics[width=18cm]{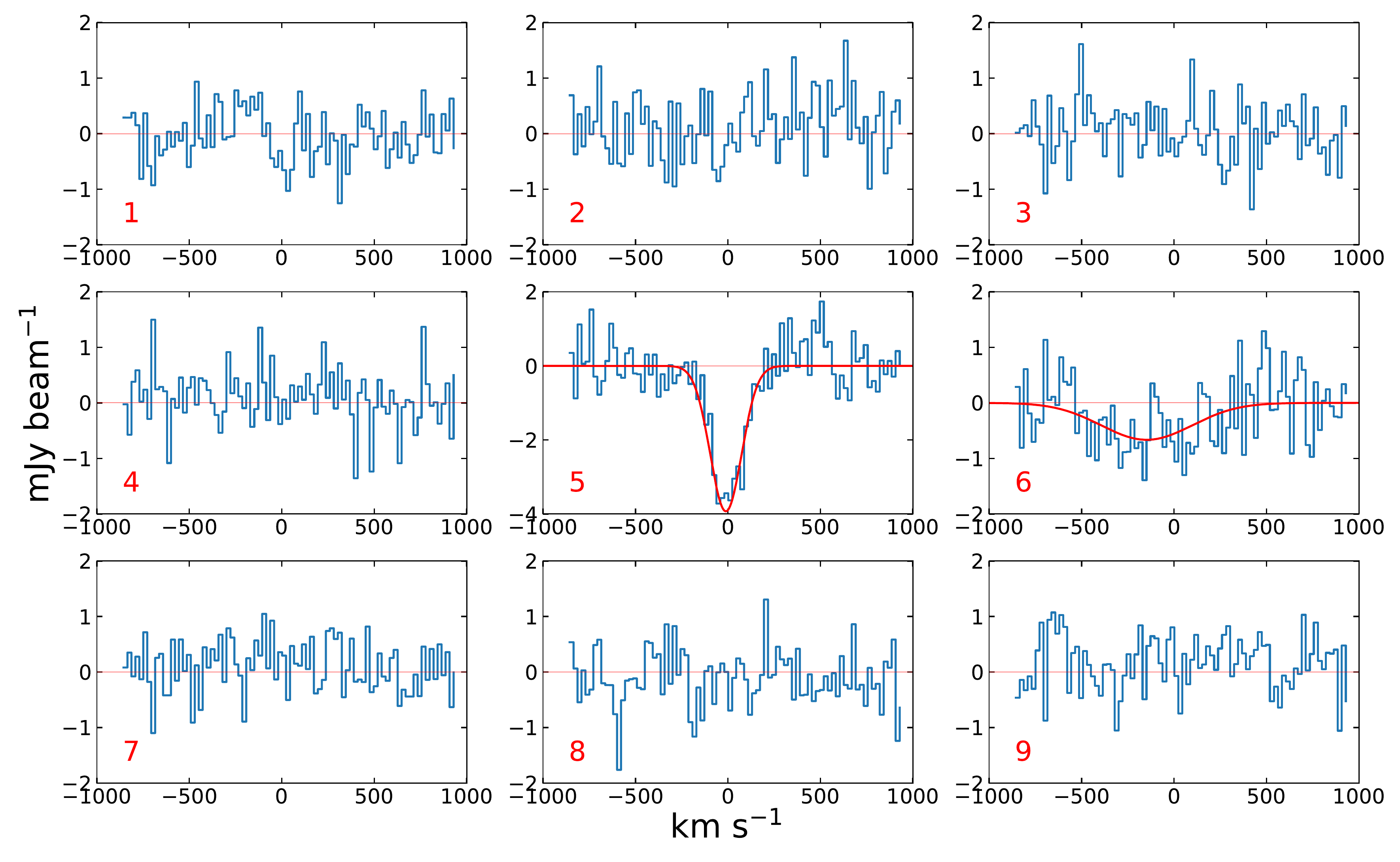}
    \caption{The extracted mean spectra from the uniform-weighting cube over each of the 9 regions shown in Fig. \ref{fig:radio_image}. We did not detect any  \mbox{H\,{\sc i}} emission, but \mbox{H\,{\sc i}} absorption is detected in squares 5 and 6. The abscissa of each plot is the rest-frame velocity relative to the systemic velocity of IRAS 10565+2448. The absorption lines seen in squares 5 and 6 are best fitted with a single Gaussian profile, see the spectra and the fitted parameters in Tab. \ref{tab:spectra_fitting}. }
    \label{fig:spectra_u}
\end{figure*}

We have successfully recovered the \mbox{H\,{\sc i}} absorption seen in the Arecibo observation, which has two prominent features. One is narrow and deep, while the another is broad and shallow \citep{mirabel1988}. To give an overview of the spectra against background radio emission, we have distributed nine red squares to cover the radio emission, see Fig. \ref{fig:radio_image}, over which we then extracted the mean spectra from uniform-weighting cube which are presented in Fig. \ref{fig:spectra_u}.

We only found \mbox{H\,{\sc i}} absorption from the central square 5 and 6, see Fig. \ref{fig:spectra_u}. What is intriguing is that we spatially resolved the two \mbox{H\,{\sc i}} absorption features, as the narrow and deep feature comes from square 5 and the broad and shallow feature comes from square 6, although our resolution can not widely separate them. 

To better illustrate the positions of the two \mbox{H\,{\sc i}} gas, we made integrated intensity maps. For the narrow and deep feature, we used a velocity range from -250 to 250 km\,s$^{-1}$ as can be seen in the mean spectrum from square 5. We found that the \mbox{H\,{\sc i}} gas comes from the central part, see the left plot in Fig. \ref{fig:gas_distri}. For the broad and shallow feature which extends from $\sim$ -500 km\,s$^{-1}$ to $\sim$ 375 km\,s$^{-1}$, we used velocity spanning from -500 to -250 km\,s$^{-1}$. The integral intensity map is shown in the middle plot in Fig. \ref{fig:gas_distri}, with the position of the blueshifted, broad, and shallow component marked with the red square 15. 

Similarly, we distributed four squares to extract spectra from the natural-weighting cube that has lower spectral noise, see the positions of the four green squares in Fig. \ref{fig:radio_image}. These spectra are shown in Fig. \ref{fig:spectra_n}. We found a possible redshifted, broad, and shallow \mbox{H\,{\sc i}} absorption line with velocity up to $\sim 500\,\rm {km\,s^{-1}}$ in the spectrum from square 10. We then made an integral intensity map using velocities from 250 to 500 km\,s$^{-1}$ and the natural-weighting cube, which is presented in the right plot in Fig. \ref{fig:gas_distri}. The location of the possible redshifted and broad \mbox{H\,{\sc i}} absorption is marked with the red square 16. Again, we found a few pixels showing positive intensity marked with the red square 17, see the section \ref{emission}.

\begin{figure*}
	\includegraphics[width=\columnwidth]{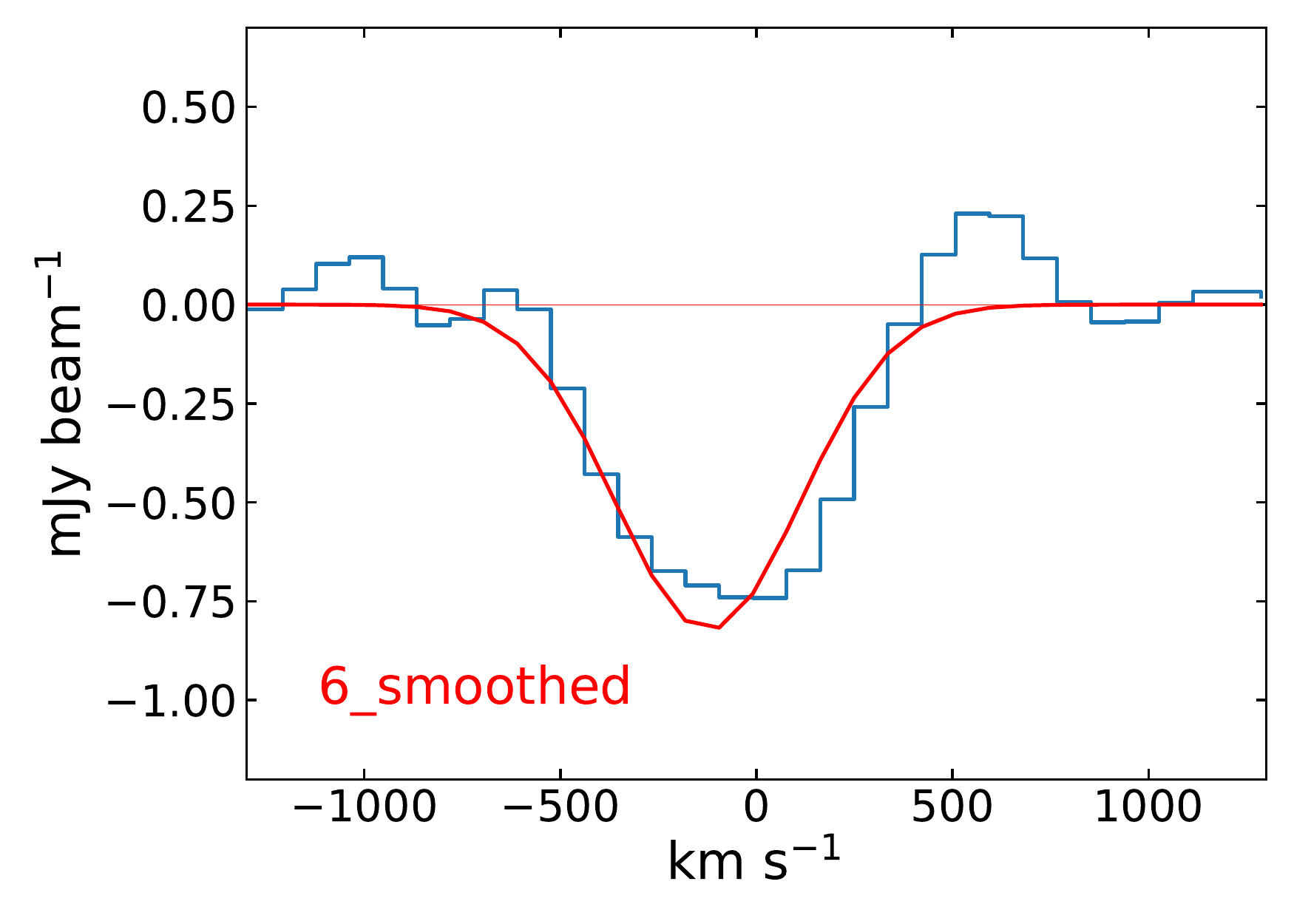}\includegraphics[width=\columnwidth]{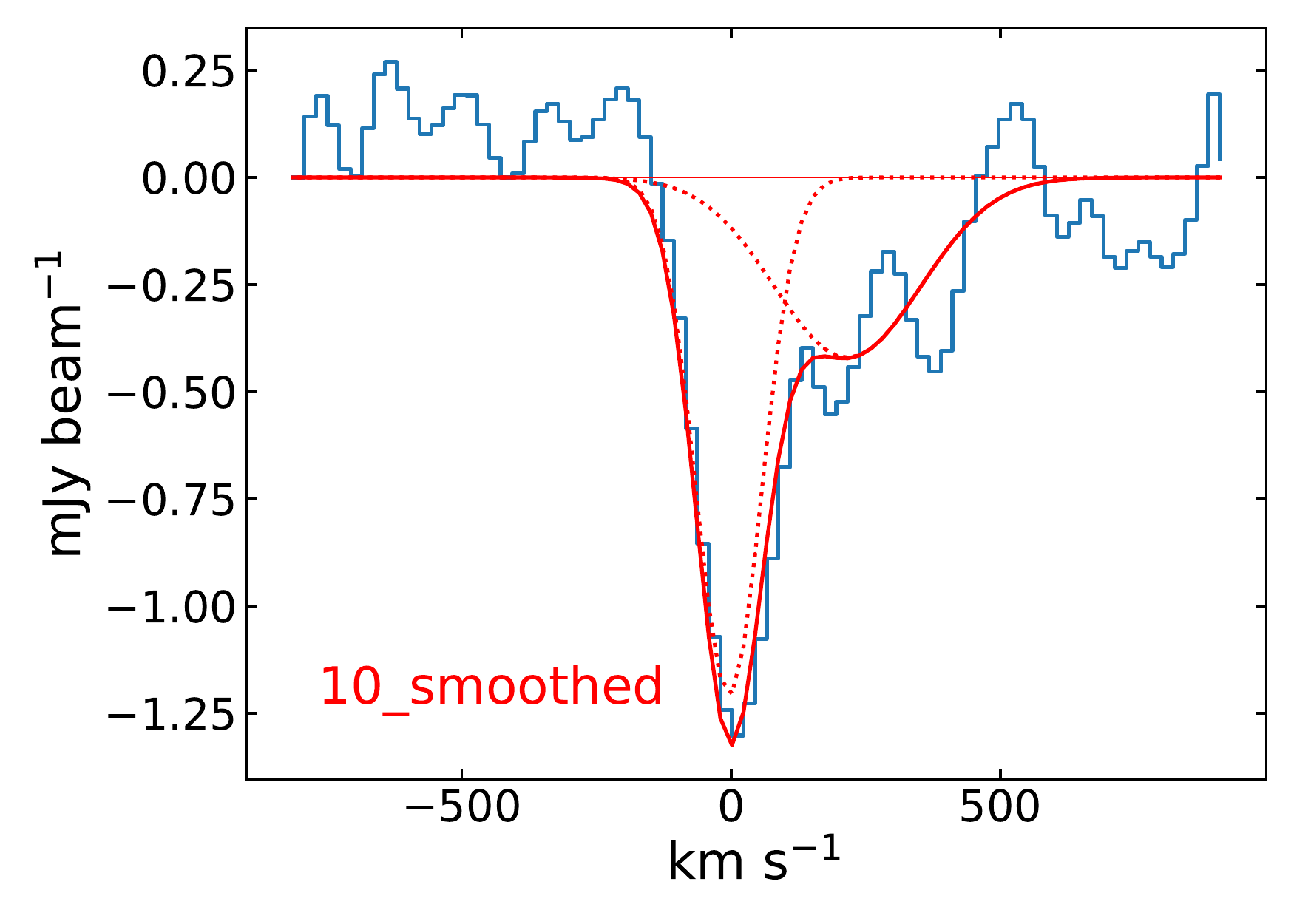}
    \caption{Left: the smoothed spectrum from square 6 in Fig. \ref{fig:spectra_u}. This spectrum was made from two steps: 1) binning every four channels of the spectrum from the square 6 to one channel; 2) Hanning smoothing over five channels. The fitted Gaussian line is shown in red. Right: the smoothed spectrum from square 10 in Fig. \ref{fig:spectra_n}. This spectrum was made by applying Hanning smooth over five channels. We fitted the spectrum with two components including a relatively faint, redshifted and broad absorption that needs to be confirmed in more sensitive observations.}
    \label{fig:smoothed_spectra}
\end{figure*}

\begin{figure*}
	\includegraphics[width=6cm]{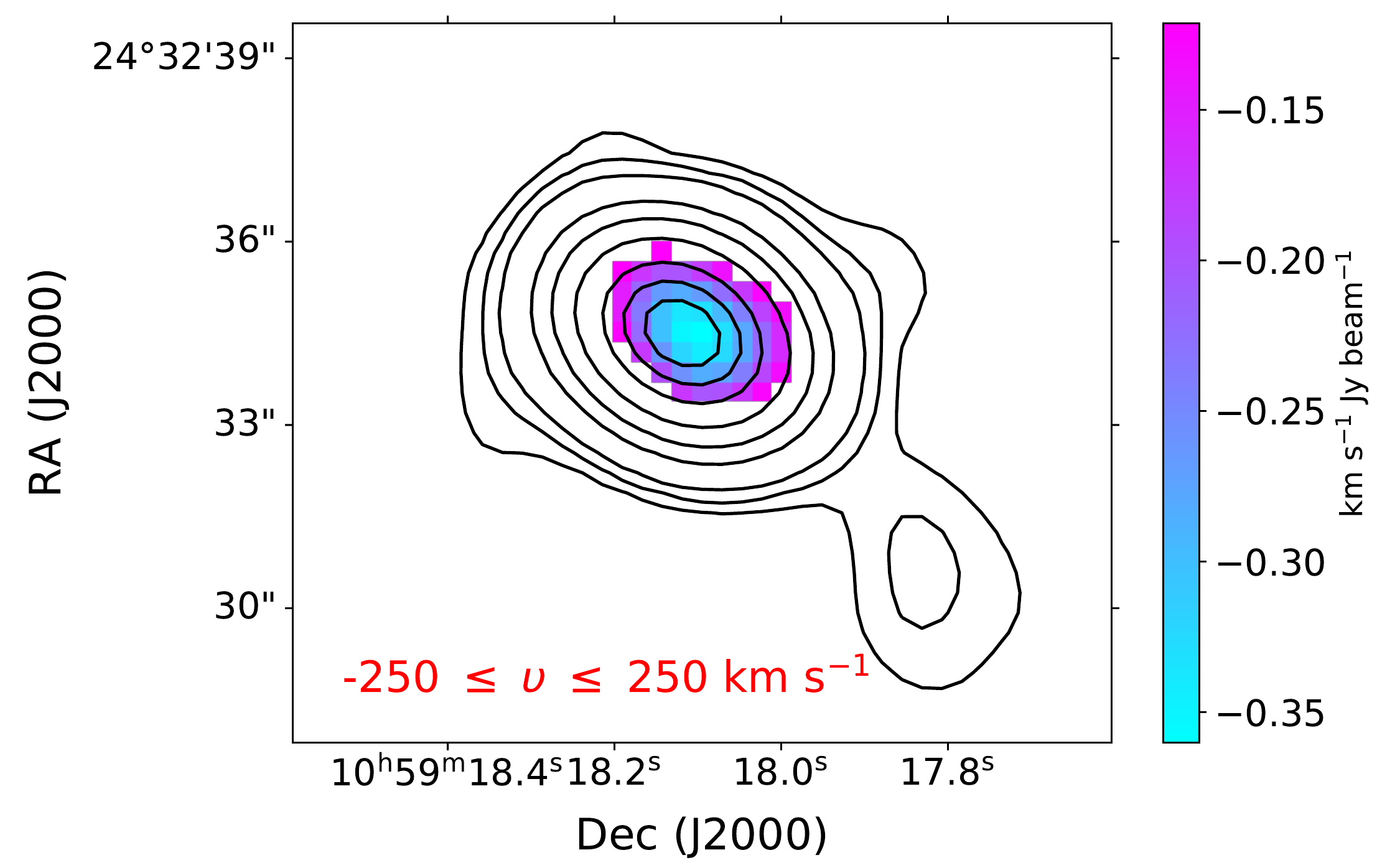}\includegraphics[width=6cm]{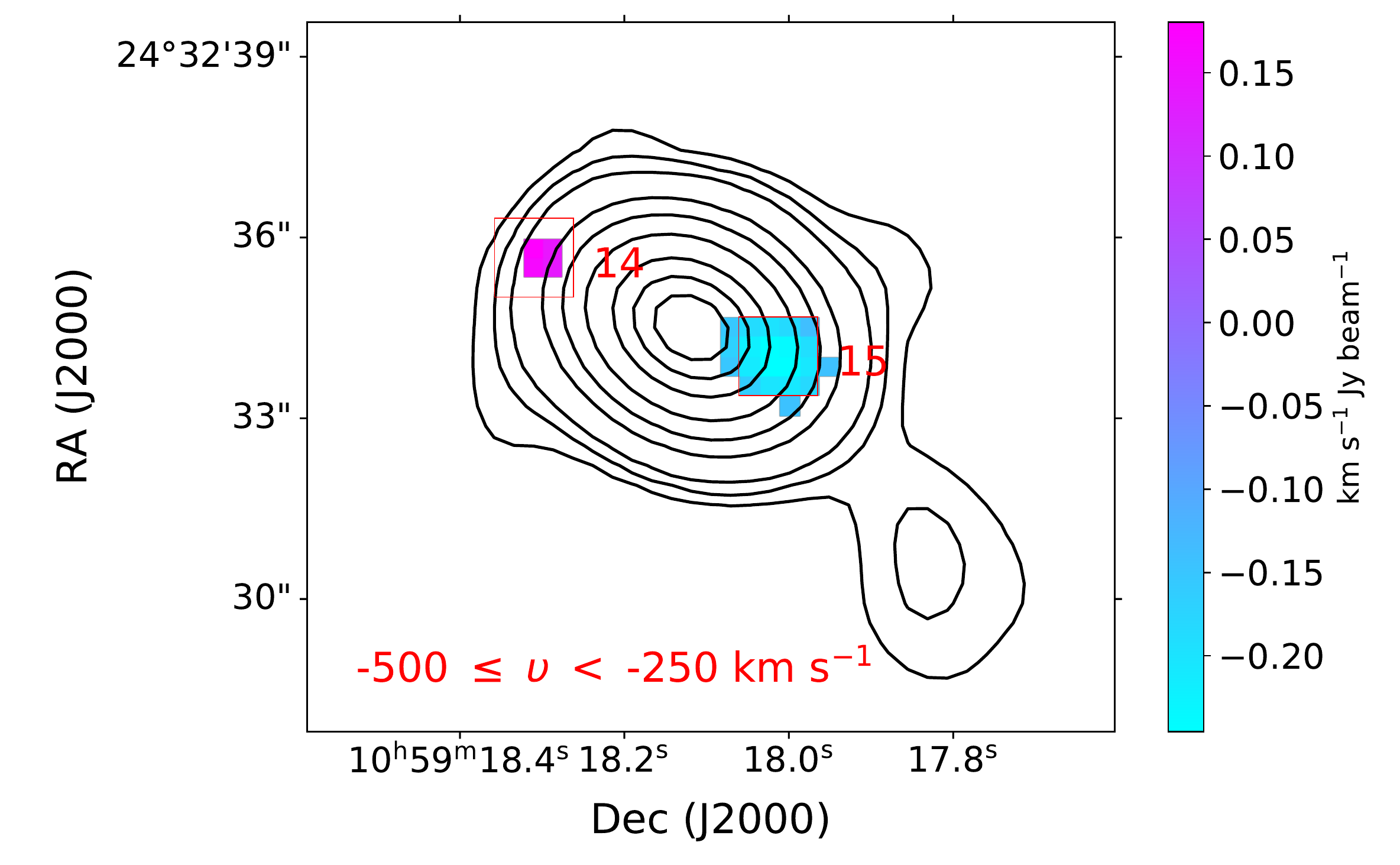}\includegraphics[width=6cm]{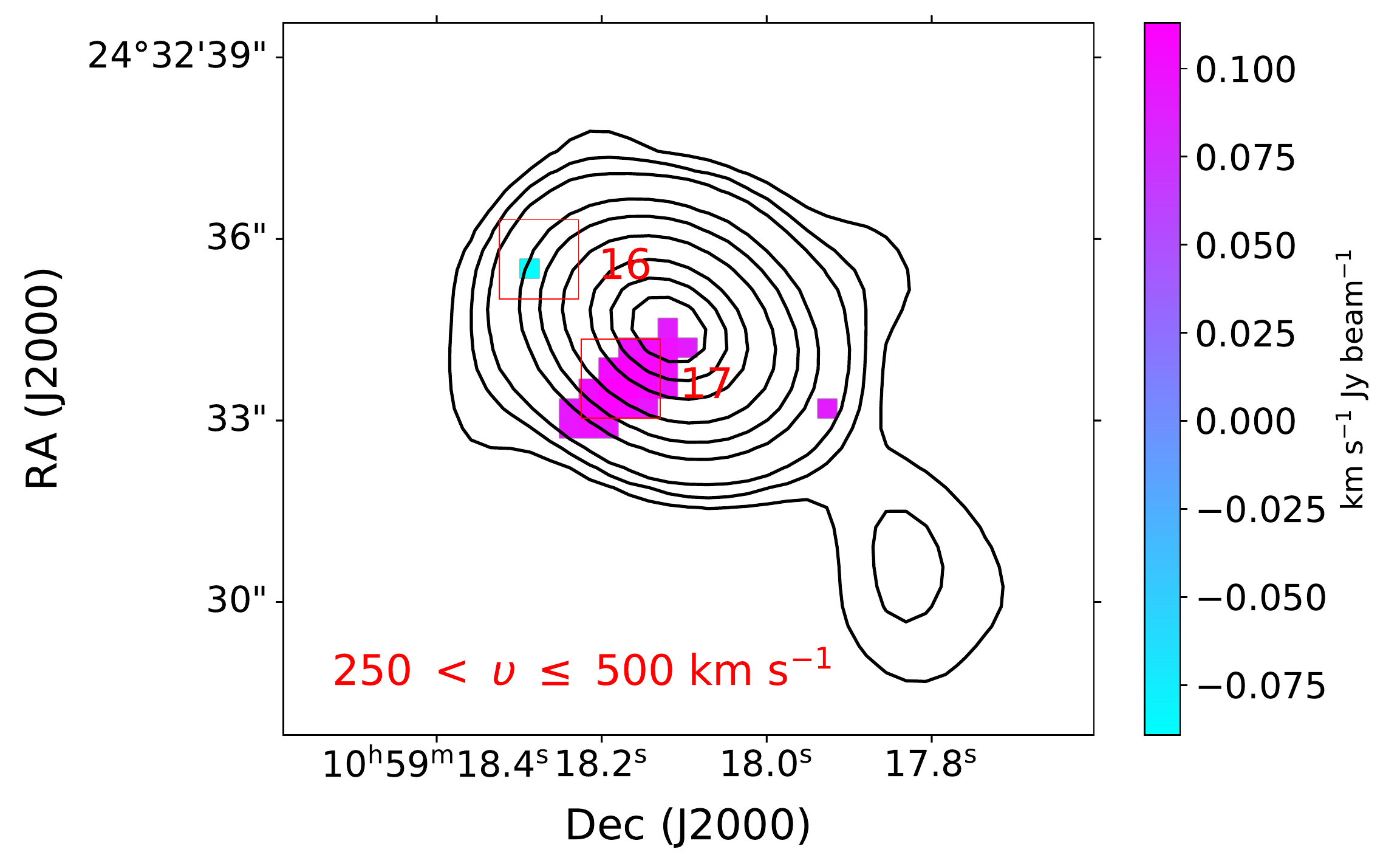}
    \caption{The distribution of integrated \mbox{H\,{\sc i}} line strength against background radio emission. Left: spanning velocity from -250 km\,s$^{-1}$ to 250 km\,s$^{-1}$ in the rest frame, using the uniform-weighting cube. Middle: spanning velocity from -500 km\,s$^{-1}$ to -250 km\,s$^{-1}$ in the rest frame, using the uniform-weighting cube. Right: spanning velocity from 250 km\,s$^{-1}$ to 500 km\,s$^{-1}$ in the rest frame, using however the natural-weighting cube. Please note that all background radio contours in these plots are the same, and made with uniform weighting. Only pixels with absolute value at least 2$\sigma$ are coloured and marked with red squares. The square 15 and 16 mark the positions of blueshifted and possible redshifted outflows. Please note that the positions of the outflows are somewhat uncertain due to the low signal-to-noise ratio. The weak \mbox{H\,{\sc i}} seen in squares 14 and 17 is likely to be spurious. It may arise from noise, or from \mbox{H\,{\sc i}} emission where the signal is too faint to be verified based on the mean wide-velocity spectrum extracted from the square 14 and 17. We show the mean spectrum from square 17 in Fig. \ref{fig:faint_HI_emi} for your reference.}
    \label{fig:gas_distri}
\end{figure*}

\begin{figure}
	\includegraphics[width=\columnwidth]{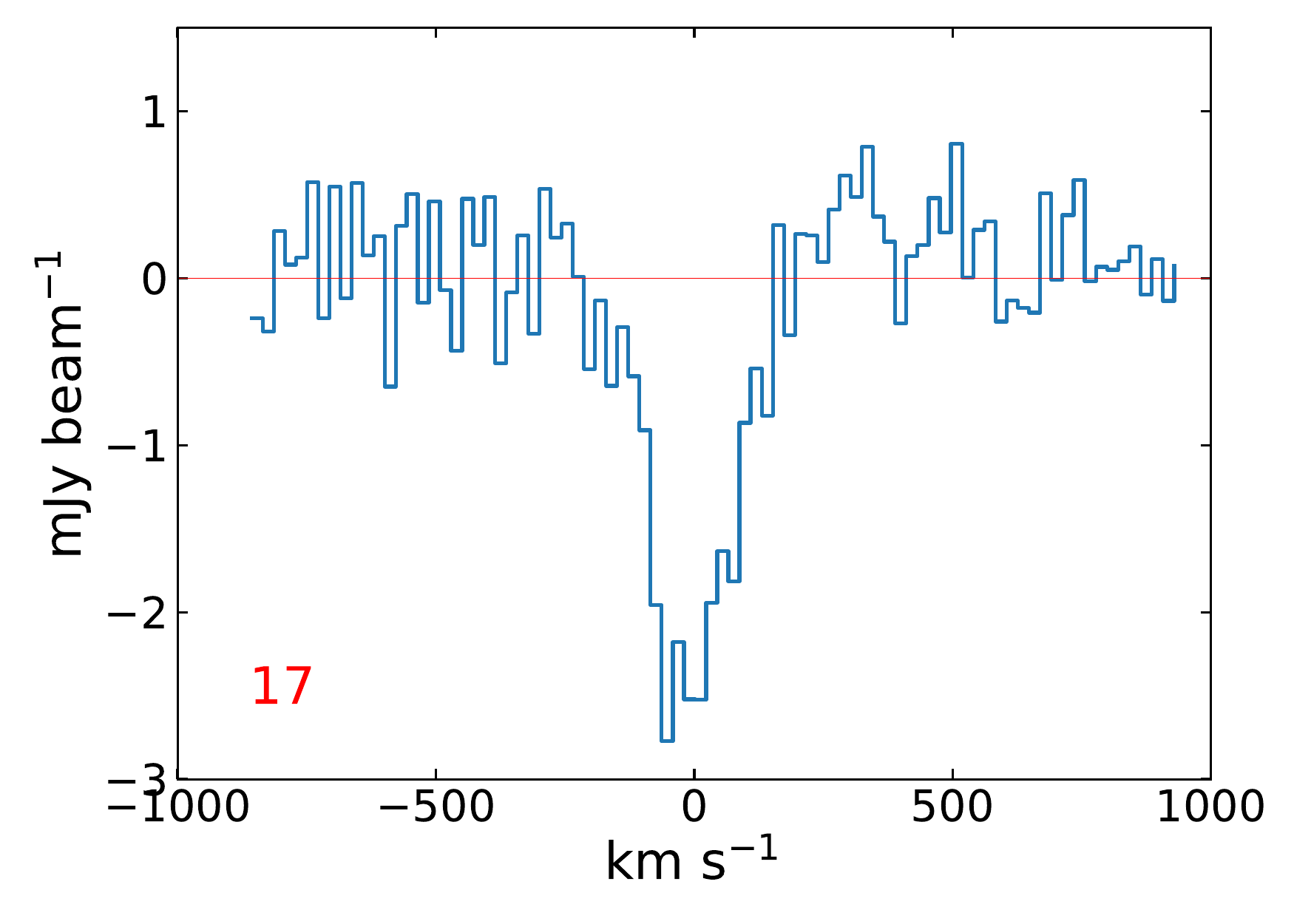}
    \caption{The mean wide-velocity spectrum extracted from pixels surrounding square 17 that have positive values of at least 2$\sigma$ in the right plot of Fig. \ref{fig:gas_distri}, using the natural-weighting cube. The emission between 250 km\,s$^{-1}$ to 500 km\,s$^{-1}$ is very faint, and may not be statistically significant. }
    \label{fig:faint_HI_emi}
\end{figure}

\begin{figure*}
	\includegraphics[width=15cm]{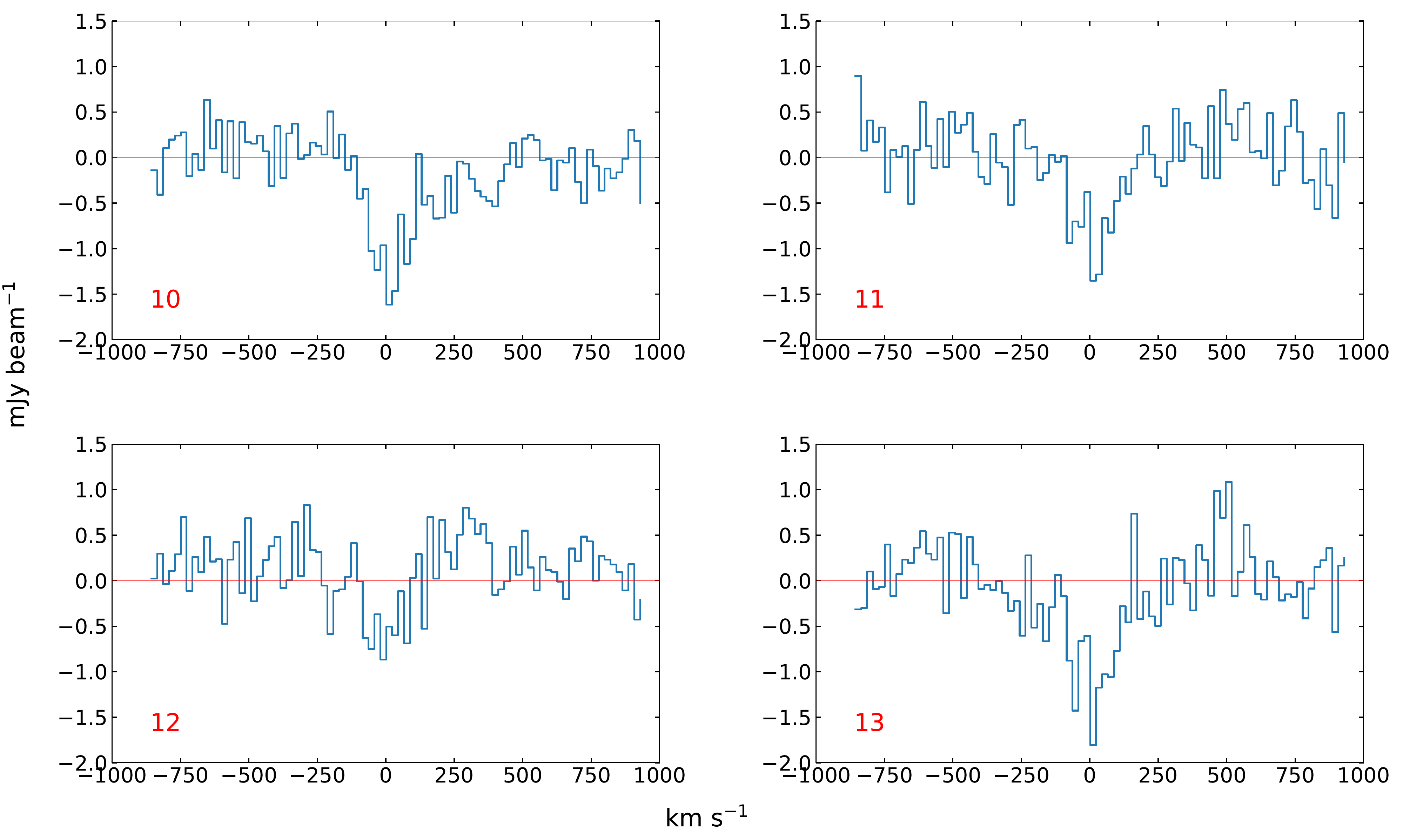}
    \caption{The extracted mean spectra from the natural-weighting cube over the 4 green squares in Fig. \ref{fig:radio_image}. The abscissa is the rest-frame velocity relative to the galaxy's systemic velocity. }
    \label{fig:spectra_n}
\end{figure*}

\begin{figure*}
	\includegraphics[width=15cm]{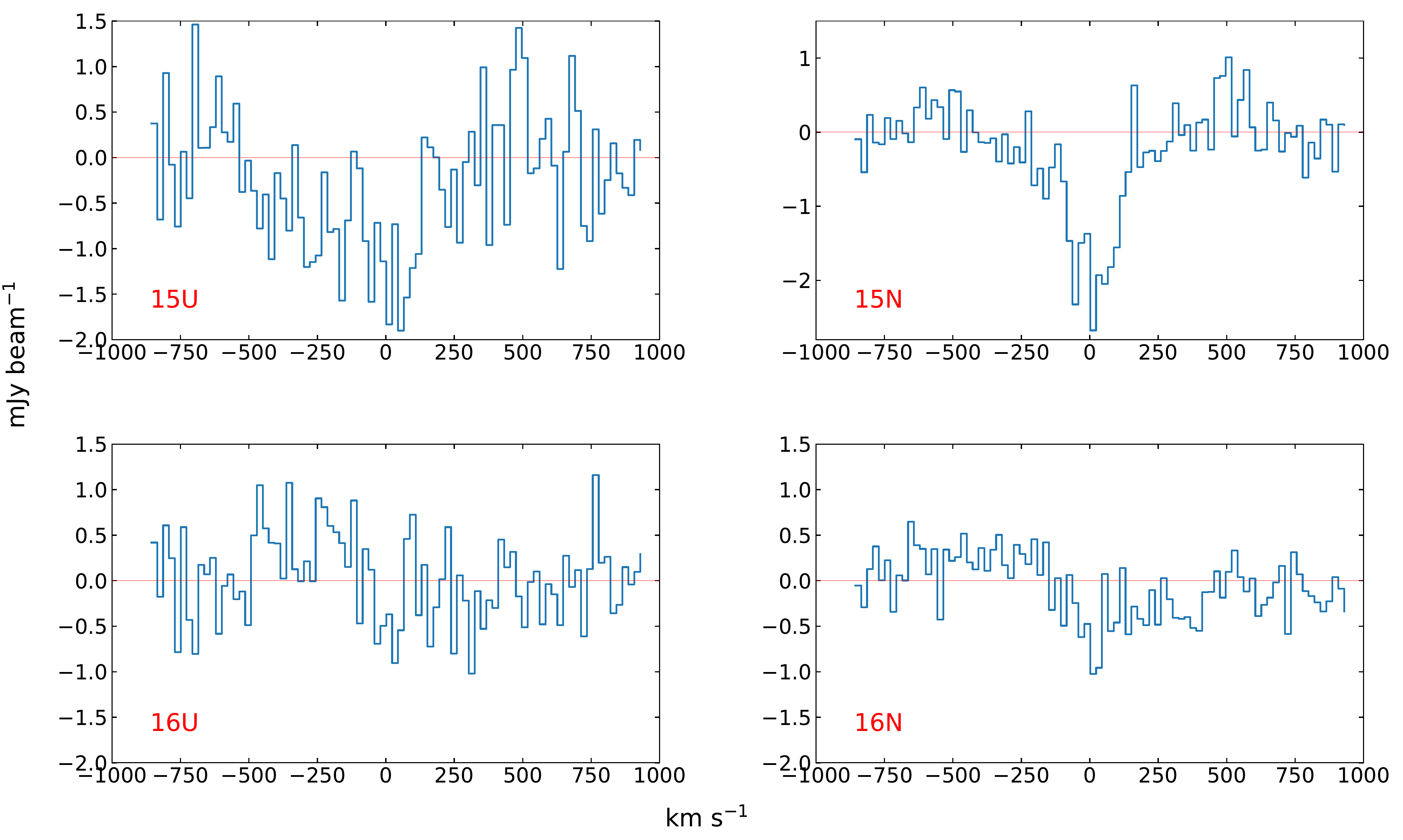}
    \caption{Upper-left: the mean spectrum extracted from square 15 using the uniform-weighting cube. Upper-right: as upper-left but using the natural-weighting cube. Lower-left: the mean spectrum extracted from square 16 using the uniform-weighting cube. Lower-right: as lower-left but using the natural-weighting cube. }
    \label{fig:spectra_outflow}
\end{figure*}

To give a further view of both the blueshifted and possible redshifted, broad, and shallow \mbox{H\,{\sc i}} absorption spectra, we also extracted the mean spectra over squares 15 and 16 from both uniform-weighting and natural-weighting cubes. We note that the positions of the outflows are somewhat uncertain due to the low signal-to-noise ratio. The extracted spectra are shown in Fig. \ref{fig:spectra_outflow}. The possible redshifted, broad, and shallow \mbox{H\,{\sc i}} absorption feature seen in the spectrum extracted over square 16 from the natural-weighting cube is more evident than that in the spectrum extracted over square 10. Further we smoothed the spectrum from the square 10 and presented the smoothed spectrum in Fig. \ref{fig:smoothed_spectra}. The redshifted absorption has a peak signal-to-noise ratio of $\sim$ 3 in the smoothed spectrum. To further test the significance of the redshifted absorption in the smoothed spectrum, we calculated the reduced $\chi^{2}$ and found two-component fitting gives a reduced $\chi^{2}$ of 1.16 while one-component fitting, however, gives a 10\% smaller value. This slight difference in reduced $\chi^{2}$ indicates that the redshifted absorption is not significant if it is real. We thus state that the detection of a redshifted, broad, and shallow \mbox{H\,{\sc i}} absorption is tentative and leave it for future more sensitive observations.    

To parametrize the \mbox{H\,{\sc i}} absorption lines we detected, we used the Bayesian method developed by \cite{allison2012} which adopts a multi-modal nested sampling, a Monte Carlo sampling algorithm, to find and fit any potential absorption line. The Bayes factor $B$ \citep{kass1995} is given to quantify the significance of the absorption lines found. $\ln(B)$ > 1 means a Gaussian spectral line model is preferred over a model with no line, see \cite{allison2012} for the details of this method. By applying this method, we obtained the parameters for the spectra from square 5 and 6 which are listed in Tab. \ref{tab:spectra_fitting}. We found both spectra are best fitted with a single Gaussian profile via a Bayesian approach as described in \cite{allison2015}. Although the signal-to-noise ratio of the spectrum from square 6 is low, the whole line has a significance of 6.9$\sigma$. We can also see the smoothed spectrum from square 6 in Fig. \ref{fig:smoothed_spectra} or the spectrum of `15U' in Fig. \ref{fig:spectra_outflow}, both showing more significant absorption and confirming the detection. 

\begin{table*}
\begin{center}
\caption{The fitted parameters for the spectra from squares 5 and 6. The numbered columns are as follows: [1] the square number, [2] the mean flux density, [3] the number of Gaussian components fitted, [4] the peak optical depth assuming a covering factor of 1, [5] the integrated optical depth, also assuming a covering factor of 1, [6] the velocity shift with respect to the systemic velocity, [7] the FWHM of the absorption line, [8] the \mbox{H\,{\sc i}} column density, where $T_{\rm s}$ is the spin temperature, 
[9] the Bayesian significance of the line, see \protect\cite{allison2012}. }
\begin{tabular}{lcccccccc}
\hline \noalign {\smallskip}
Square&$S_{v}$ & Nu.Gau. & $\tau_{\rm peak}$ & $\tau_{\rm int}$ & $v$& FWHM &$N_{\rm HI}$  & ${\rm ln}\,B$ \\

 &$\rm mJy\,beam^{-1}$ &  &  &  & km\,s$^{-1}$& km\,s$^{-1}$ &atoms\,cm$^{-2}$  &  \\
\hline \noalign {\smallskip}

5&30.07 &1&$0.14_{-0.01}^{+0.01} $ & $28.16_{-1.97}^{+1.99} $& $-11_{-6}^{+6} $ & $200_{-14}^{+16} $&$5.13_{-0.36}^{+0.36}\times10^{19}T_{\rm s}$&145\\
\\
6&9.84 &1&$0.07_{-0.02}^{+0.02}$ & $43.03_{-8.89}^{+9.05}$ & $-148_{-63}^{+62}$ &$ 581_{-103}^{+124}$ &$7.83_{-1.61}^{+1.65}\times10^{19}T_{\rm s}$&9\\

\hline \noalign {\smallskip}
\end{tabular}
\label{tab:spectra_fitting}
\medskip
\end{center}
\end{table*}

\section{Discussion}
\subsection{The gas traced by the deep and narrow \mbox{H\,{\sc i}} absorption}
The deep and narrow \mbox{H\,{\sc i}} absorption, spanning from $\sim-250\,\rm{km\,s^{-1}}$ to $\sim250\,\rm{km\,s^{-1}}$, was found only towards the square 5 that has an angular size of 1.32 arcsec $\times$ 1.32 arcsec corresponding to a linear size of 1.12 kpc $\times$ 1.12 kpc. Therefore the gas traced seems to come from a compact region. If we are tracing the neutral gas from the large galactic disc, we should also see the absorption from the surrounding squares, which is not the case here. For example, the peak optical depth of the absorption is 0.14, see Tab. \ref{tab:spectra_fitting}, which would let us see the absorption at 2$\sigma$ level from square 6 that has a mean flux density of 9.84 mJy\,beam$^{-1}$. However, despite the broad absorption from square 6, there is no other absorption. Also, the FWHM of the deep and narrow absorption is about 200 $\rm km\,s^{-1}$ which is generally much larger than those tracing large galactic disks \citep[e.g.][]{gallimore1999,serra2012,curran2016,maccagni2017}. Therefore, the best interpretation is that we are tracing the gas from the circumnuclear disc or torus. The calculated \mbox{H\,{\sc i}} column density is $5.13_{-0.36}^{+0.36}\times10^{19}T_{\rm s}$ where $T_{\rm s}$ is the spin temperature, and we assume that the background radio emission from square 5 is fully covered by the \mbox{H\,{\sc i}} gas, i.e. the covering factor is unity. 
\subsection{The gas traced by the shallow \mbox{H\,{\sc i}} absorption}
Broad and shallow \mbox{H\,{\sc i}} absorption lines are often interpreted as jet-driven outflows \citep[e.g.][]{morganti2005,mahony2013}. In our observations, we successfully recovered the blueshifted, broad and shallow \mbox{H\,{\sc i}} absorption, confirming the outflow. Besides, we further detected a possible redshifted, broad and shallow \mbox{H\,{\sc i}} absorption line.

In previous studies, redshifted \mbox{H\,{\sc i}} absorption is often interpreted as inflow \citep[e.g.][]{morganti2009,araya2010,struve2012,maccagni2014}. However, the FWHM of redshifted gas is generally narrow, around 10--20 $\rm km\,s^{-1}$, and much smaller than the FWHM, $310_{-49}^{+49}$ $\rm km\,s^{-1}$, measured from the smoothed redshifted \mbox{H\,{\sc i}} line in Fig. \ref{fig:smoothed_spectra}.

Thus it appears the possible redshifted, broad, and shallow line traces a redshifted outflow, as in principle we could detect the redshifted outflow through absorption provided the \mbox{H\,{\sc i}} gas is in front of background radio continuum. 

\subsection{Comparison with outflows detected in IRAS 10565+2448 in other bands}\label{sec:multi_phase_outflows}
Studies of the outflows in IRAS 10565+2448 have been carried out at multiple wavelengths. Optical observations of $\rm H\alpha$ emission with the Integral Field Unit (IFU) on the Gemini Multi-Object Spectrograph \cite{rupke2013} revealed widespread ionised outflows in IRAS 10565+2448. However, the $\sim 1$ kpc scale ionised outflow extending from the nucleus to southeast is different from those in other regions, reflected by the higher FWHM, see Figure 13 in \cite{rupke2013}. This indicates that there may be various kinds of outflows driven by different mechanisms. Besides, there is a larger, $\sim$ 2--3 kpc, and more diffuse ionised outflow extending to the southwest, see Figure 17 in \cite{rupke2013}. The warm ionised outflows in IRAS 10565+2448 have a mean central velocity of -133 km\,s$^{-1}$ and mean maximum velocity of -535 km\,s$^{-1}$ \citep{rupke2013}. The mass outflow rate of ionised gas was estimated as 1.4 $\rm M_\odot \,yr^{-1}$, corresponding to an energy loss rate of $8.5\times10^{40}\,\rm erg\,s^{-1}$ \citep{rupke2013}. 

\mbox{Na\,{\sc i}} D absorption, which is often used as a tracer of neutral hydrogen, was also detected in IRAS 10565+2448 with a column density of $3.0\times10^{21}$ atoms\,cm$^{-2}$ in long-slit spectroscopic observations \citep{rupke2005a,rupke2005}, which is about half of that detected in the our blueshifted \mbox{H\,{\sc i}} outflow if adopting $T_{\rm s}=100$ K. Follow up IFU observations spatially resolved the \mbox{Na\,{\sc i}} D absorption and found widespread blueshifted \mbox{Na\,{\sc i}} D absorption indicating outflow in IRAS 10565+2448 \citep{rupke2013}. Like the ionised outflow, the \mbox{Na\,{\sc i}} D outflow has a similar spatial distribution with a diffuse component extending to the southwest, see Figure 17 in \cite{rupke2013}. \mbox{Na\,{\sc i}} D outflows in IRAS 10565+2448 have a mean central velocity of -218 km\,s$^{-1}$ and mean maximum velocity of -468 km\,s$^{-1}$ \citep{rupke2013}. The mass outflow rate of neutral gas estimated from \mbox{Na\,{\sc i}} D absorption is 64.6 $\rm M_\odot \,yr^{-1}$, corresponding to an energy loss rate of $2.6\times10^{42}\,\rm erg\,s^{-1}$ \citep{rupke2013}.

More importantly, IRAM PdBI observations detected both blueshifted and redshifted cold molecular CO outflows, revealed as broad emission wings spanning velocities -300 km\,s$^{-1}$ $\sim$ -600 km\,s$^{-1}$ and 300 km\,s$^{-1}$ $\sim$ 600 km\,s$^{-1}$ \citep{cicone2014}. The image of the CO outflows shows that the blueshifted outflow extends from the nucleus to the southwest, whereas the redshifted outflow extends from the nucleus to the northeast with an estimated radius of $\sim$ 1.1 kpc \citep{cicone2014}. Assuming a spherical geometry and adopting a CO-to-$\rm H_{\rm2}$ conversion factor of 0.8 $M_\odot/(\rm K\,km\,s^{-1}\,pc^{2})$ that is commonly used for the molecular ISM of ULIRGs \citep{bolatto2013}, a $\rm H_{\rm2}$ gas mass outflow rate of 300 $\rm M_\odot \,yr^{-1}$ was obtained. \cite{fluetsch2019} studied the cold molecular outflows in several nearby galaxies, including IRAS 10565+2448. By assuming the outflow is expelled shell-like, the $\rm H_{\rm2}$ gas mass outflow rate was estimated as 100 $\rm M_\odot \,yr^{-1}$, corresponding to an energy loss rate of $6.3\times10^{42}\,\rm erg\,s^{-1}$. Please note that the $\rm H_{2}$ gas considered here only comes from wings spanning velocity -300 km\,s$^{-1}$ $\sim$ -600 km\,s$^{-1}$ and 300 km\,s$^{-1}$ $\sim$ 600 km\,s$^{-1}$. By using the data from \cite{rupke2013} and \cite{cicone2014}, \cite{fluetsch2021} studied this source again. Interestingly, blueshifted OH 199 $\mu$m absorption was detected in IRAS 10565+2448 \citep{veilleux2013}, giving a mass outflow rate of 248 $\rm M_\odot \,yr^{-1}$ \citep{gonzalez2017}.

Our observations have detected the blueshifted and possible redshifted neutral hydrogen outflows through absorption and support the model where the outflows are expelled shell-like, as the broad wing, from -500 km\,s$^{-1}$ to -250 km\,s$^{-1}$, absorption strength from square 6 is larger than that from square 5 where the radio emission is brightest and so should have stronger absorption if the \mbox{H\,{\sc i}} column density is uniform across squares 5 and 6, see also the positions of the outflows in Fig. \ref{fig:gas_distri}. From our observation, the blueshifted and possible redshifted \mbox{H\,{\sc i}} outflows are quite symmetric with respect to the nucleus. To calculate the \mbox{H\,{\sc i}} mass outflow rate, we followed the equation \citep{heckman2002}: 
\begin{equation}\label{mass_outflow_rate}
\dot{M} = 30 \frac{\Omega}{4\pi} \frac{r_*}{1\,\mathrm{kpc}} \frac{N_{\mathrm{HI}}}{10^{21}\mathrm{cm}^{-2}} \frac{v}{300\,\mathrm{km\,s}^{-1}} \rm M_\odot\,{\rm yr}^{-1}
\end{equation}

\noindent where the $\Omega$ is the solid angle of outflow, $r_*$ is the outflow radius, $N_{\mathrm{HI}}$ is the \mbox{H\,{\sc i}} column density, and $v$ is the outflow velocity. 

Since the detection of redshifted outflow is not significant, we temporarily exclude the redshfited outflow from our calculation of outflow properties.

We use the spectrum from square 6 as the blueshifted outflow spectrum, and the distance from the radio nucleus to the centre of square 6 as the outflow radius. We chose not to use the spectrum from square 15 because it seems slightly influenced by the narrow and deep absorption from square 5, see Fig. \ref{fig:spectra_outflow}, whereas the spectrum from square 6 looks unpolluted. Actually, square 6 is almost overlapped with square 15 and so using the spectrum and position information from either square 6 or 15 would not make a big difference. 

Using the spectrum from square 6, we found that the blueshifted outflow has a velocity up to $\sim$ -530 km\,s$^{-1}$ with the central velocity at $\sim$ -148 km\,s$^{-1}$. The corresponding \mbox{H\,{\sc i}} column density is $\sim 7.83\times10^{19}T_{\rm s}$ atoms\,cm$^{-2}$ and the projected distance between the radio nucleus and the centre of square 6 is 1.36 kpc, which was used as the outflow radius. By assuming the blueshifted outflow has a solid angle of $\pi$, we derive a mass outflow rate of $\sim0.39T_{\rm s}$ $\rm M_\odot \,yr^{-1}$ for the blueshifted outflow. 

We can calculate the energy loss rate by using the equation \citep{holt2006}:
\begin{equation}\label{energy_loss_rate}
\dot{E} = 6.34 \times10^{35} \frac{\dot{M}}{2} \left(v^2 + \frac{{\rm FWHM}^2}{1.85}\right) {\rm erg}\,{\rm s}^{-1}, 
\end{equation}

\noindent where FWHM is the full width at half maximum, which gives a energy loss rate of $\sim 2.6\times10^{40}T_{\rm s}$\ $\rm erg\,s^{-1}$ for the blueshifted \mbox{H\,{\sc i}} outflow in IRAS 10565+2448.

The larger, $\sim$ 2--3 kpc, more diffuse, and warm ionised outflow detected through optical $\rm H\alpha$ emission and diffuse neutral outflow extending to southwest detected through \mbox{Na\,{\sc i}} D absorption, the cold molecular outflows detected through CO emission, and the neutral hydrogen outflow detected through \mbox{H\,{\sc i}} absorption all are aligned from northeast to southwest, indicating they may be multi-phase gas counterparts of the same outflow. 

To more precisely understand the impact of outflows in IRAS 10565+2448, we need to combine the results from these observations. For the ionised outflow, we directly quote the mass outflow rate and energy loss rate from $\rm H\alpha$ gas from \cite{rupke2013}, which is negligible compared to the molecular and neutral outflows. For the neutral hydrogen outflow, we use the results of our own observations. We obtained a mass outflow rate of $\sim0.39T_{\rm s}$ $\rm M_\odot \,yr^{-1}$ corresponding to an energy loss rate of $\sim 2.6\times10^{40}T_{\rm s}$\ $\rm erg\,s^{-1}$. Lastly, we use the calculations from CO emission in \cite{fluetsch2019} for the molecular $\rm H_2$ outflow. The reason of why we don't use the OH 199 $\mu$m absorption from \cite{veilleux2013} is that absorption can only reveal foreground gas. We don't use the CO analysis in \cite{cicone2014} because spherical geometry was assumed for the outflow whereas our uGMRT observations support expelled shell geometry that used by \cite{fluetsch2019}. A mass outflow rate of $100\,\rm M_\odot \,yr^{-1}$ and an energy loss rate of $\sim 6.3\times10^{42}\,\rm erg\,s^{-1}$ were given to the $\rm H_2$ gas outflow in \cite{fluetsch2019}. The properties of these outflows are summarised in Tab. \ref{tab:outflows_properties}.

\begin{table}
\begin{center}
\caption{The properties of multi-phase gas outflows in IRAS 10565+2448. Columns 1--5 list: the gas tracer, radius, velocity, mass outflow rate, and energy loss rate. For the ionised gas, we directly quote the values from \protect\cite{rupke2013}. For the H$_2$ gas, we directly adopt the values from \protect\cite{fluetsch2019}. Please note that the H$_2$ gas properties were calculated by only using the high velocity, $|v|>300\,\rm km\,s^{-1}$, part of CO emission.}
\begin{tabular}{lcccc}
\hline \noalign {\smallskip}

&$r_{*}$ &$v$ & $\dot{M}$ & $\dot{E}$ \\

&kpc &$\rm km\,s^{-1}$ & $M_\odot\,{\rm yr}^{-1}$ &$\rm erg\,s^{-1}$ \\

\hline \noalign {\smallskip}

$\rm H\alpha$ &1 &-133 & 1.4 & $8.5\times10^{40}$ \\

$\rm H_2$ &1.1 &-450 & 100 & $6.3\times10^{42}$\\

$\rm HI$ &1.36 &-148 & $0.39T_{\rm s}$ & $2.6\times10^{40}T_{\rm s}$ \\

\hline \noalign {\smallskip}
\end{tabular}
\label{tab:outflows_properties}
\medskip

\end{center}
\end{table}

The \mbox{H\,{\sc i}} gas spin temperature is a key factor in estimating the impact of the outflow. Previously known measurements of \mbox{H\,{\sc i}} spin temperature imply that 100\,K should be taken as a lower limit \citep[e.g.][]{reeves2015,reeves2016,kanekar2014,murray2018,allison2021}. If we assume an \mbox{H\,{\sc i}} spin temperature of 100\,K in IRAS 10565+2448, then the total mass outflow rate is $\sim 140\, \rm M_\odot \,yr^{-1}$, corresponding to a total energy loss rate $\sim  8.9\times10^{42}\,\rm erg\,s^{-1}$.

Please note that $\sim 140\, \rm M_\odot \,yr^{-1}$ and $\sim  8.9\times10^{42}\,\rm erg\,s^{-1}$ should be taken as lower limits due to the reasons below: 1), the low velocity part, $|v|\leq300\,\rm km\,s^{-1}$, of CO emission were not taken into consideration; 2), the possible redshifted \mbox{H\,{\sc i}} outflow has been excluded and the \mbox{H\,{\sc i}} spin temperature we used is 100 K which is at the lower end of the range of observed $T_{\rm s}$, especially given the presence of an ionised outflow counterpart; 3), we have not corrected the projection effect because our observed quantities including radial velocity and FWHM are projected value. 

\subsection{What drives the outflow in IRAS 10565+2448?}
Both starbursts and AGN can drive outflows. Although the star formation rate in IRAS 10565+2448 is as high as 131.8 $\rm M_\odot$\,year$^{-1}$ \citep{u2012}, our analysis suggests that the radio jet may play a role in driving outflows. 

\subsubsection{Starburst-driven outflow?}
In star-forming galaxies, the kinetic power released by supernovae (SNe) is capable of driving outflows. \cite{rupke2013} concluded that a starburst is the dominant source of driving outflows in IRAS 10565+2448 because there is no sign of AGN from optical and X-ray bands \citep{veilleux1995,yuan2010,teng2010} or the AGN is weak whose bolometric luminosity only occupies 17 percent of the total bolometric luminosity of the galaxy based on the polycyclic aromatic hydrocarbon (PAH) strength and mid-infrared (MIR) spectral shape \citep{veilleux2009}.  However, there is no specific calculation in \cite{rupke2013} to test whether a starburst can drive the observed outflows.  

Here, we take a cautious approach. Following \cite{veilleux2005} we estimated the mass outflow rate driven by SNe as $\dot{M} = 0.26({\rm SFR}/{\rm M_\odot \, yr^{-1})}$\,${\rm M_\odot \, yr^{-1}}$, which is $\sim 34.3\,{\rm M_\odot \, yr^{-1}}$ -- much lower than the value of $\sim 140\,\rm M_\odot \,yr^{-1}$ that we have observed in IRAS 10565+2448. Since the mass flow rate derived from the models only considers the mass flux from the hot wind produced by the starburst, the actual outflow rate could be several times higher if (as expected) this wind entrains mass from the host galaxy. If sufficient additional mass is entrained, then the outflow rate could plausibly reach the observed value of $\sim 140\,\rm M_\odot \,yr^{-1}$. 


From the kinetic power view, the kinetic power associated with the outflows is at least $8.9\times10^{42}\,\rm erg\,s^{-1}$ whereas SNe can release, $P_{\rm K,SF}=7.0\times10^{41}\,(SFR/{\rm M_\odot \, yr^{-1}})$\,${\rm erg\,s^{-1}}$, $\sim 6.7\times10^{43}\,\rm erg\,s^{-1}$. If supernovae are the main driver for the outflows, then the thermalization efficiency, the fraction of SNe kinetic power converted to outflow kinetic power, needs to be at least $\sim 0.13$. 

IRAS 10565+2448 is classified as a `composite' galaxy based on the $\rm [OIII]5007/H\beta \,vs\,[NII]6583/H\alpha$ diagram and an `HII' galaxy based on $\rm [OIII]5007/H\beta \,vs\,([SII]6716+6731)/H\alpha$ and $\rm [OIII]5007/H\beta \,vs\,[OI]6300/H\alpha$ diagrams \citep{yuan2010}. Using the spectral classification scheme presented in \cite{veilleux1987}, IRAS 10565+2448 is classified as `HII' as well \citep{yuan2010}. 
Thus, IRAS 10565+2448 seems most likely to be an `HII' galaxy. \citep{fluetsch2019} observed a large sample of `HII' galaxies and derived a typical thermalization efficiency of about 1--2 percent. In the galaxy M82 the thermalization efficiency could be as high as 100\% \citep{strickland2009}, but this is an extreme case. 
We note that the star formation rate in M82 is only about $\sim 10\rm M_\odot \, yr^{-1}$\citep{deg2001}, which is much lower than that in IRAS 10565+2448. 
Although we can’t completely exclude that the thermalization efficiency in IRAS 10565+2448 is as high as 0.13, this would be unusually high in comparison to the star-forming systems studied by \cite{fluetsch2019}.

Besides the contribution from SNe, the radiation pressure from young starbursts can also produce outflows \citep[e.g.][]{thompson2005,thompson2015}. The momentum rate of the outflows is at least $\dot{M}V =3.3\times10^{35}\,\rm g\,cm\,s^{-2}$, while the radiation momentum rate is $\sim 1.1\times10^{35}\,\rm g\,cm\,s^{-2}$ estimated from the bolometric luminosity of the host galaxy \citep{cicone2014}. We can see that the outflow momentum rate is much higher than the radiation momentum rate, whereas from observations the ratio between outflow momentum rate and radiation momentum rate ranges from 0.1 to 0.5 \citep{fluetsch2019}. More evidence can be seen from the \mbox{H\,{\sc i}} outflow FWHM which is about 581 km\,s$^{-1}$ and much larger than the neutral outflow FWHM, $\sim$275 km\,s$^{-1}$, seen from large starbursts \citep{veilleux2005}.

As stated in Section \ref{sec:multi_phase_outflows}, there may be various outflows driven by different mechanisms. Based on the arguments above, we think it is unlikely that the starburst alone could drive the observed outflows.

\subsubsection{Evidence for an AGN in IRAS 10565+2448.}
\begin{figure*}
\includegraphics[height=\columnwidth]{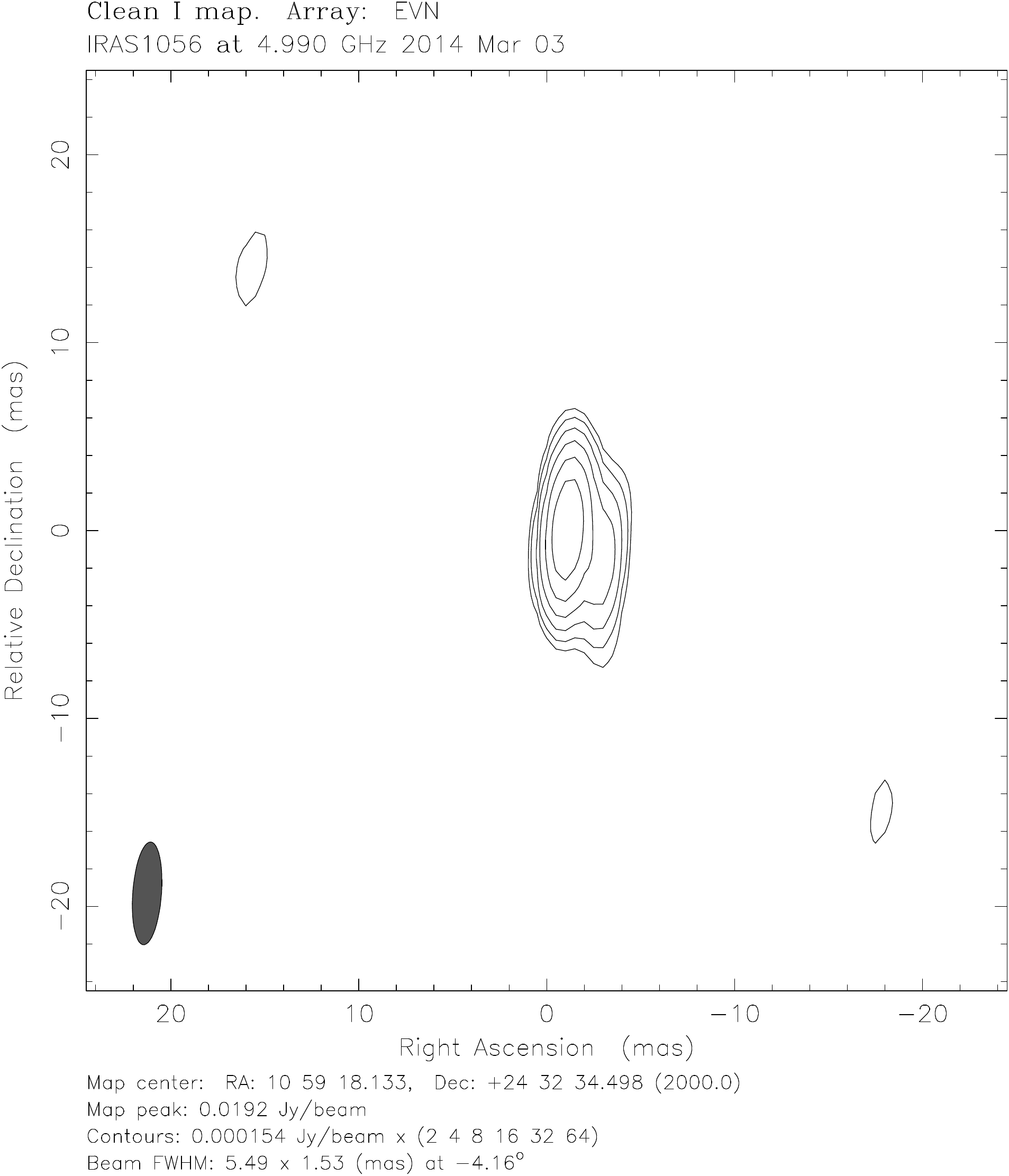} 
\includegraphics[height=\columnwidth]{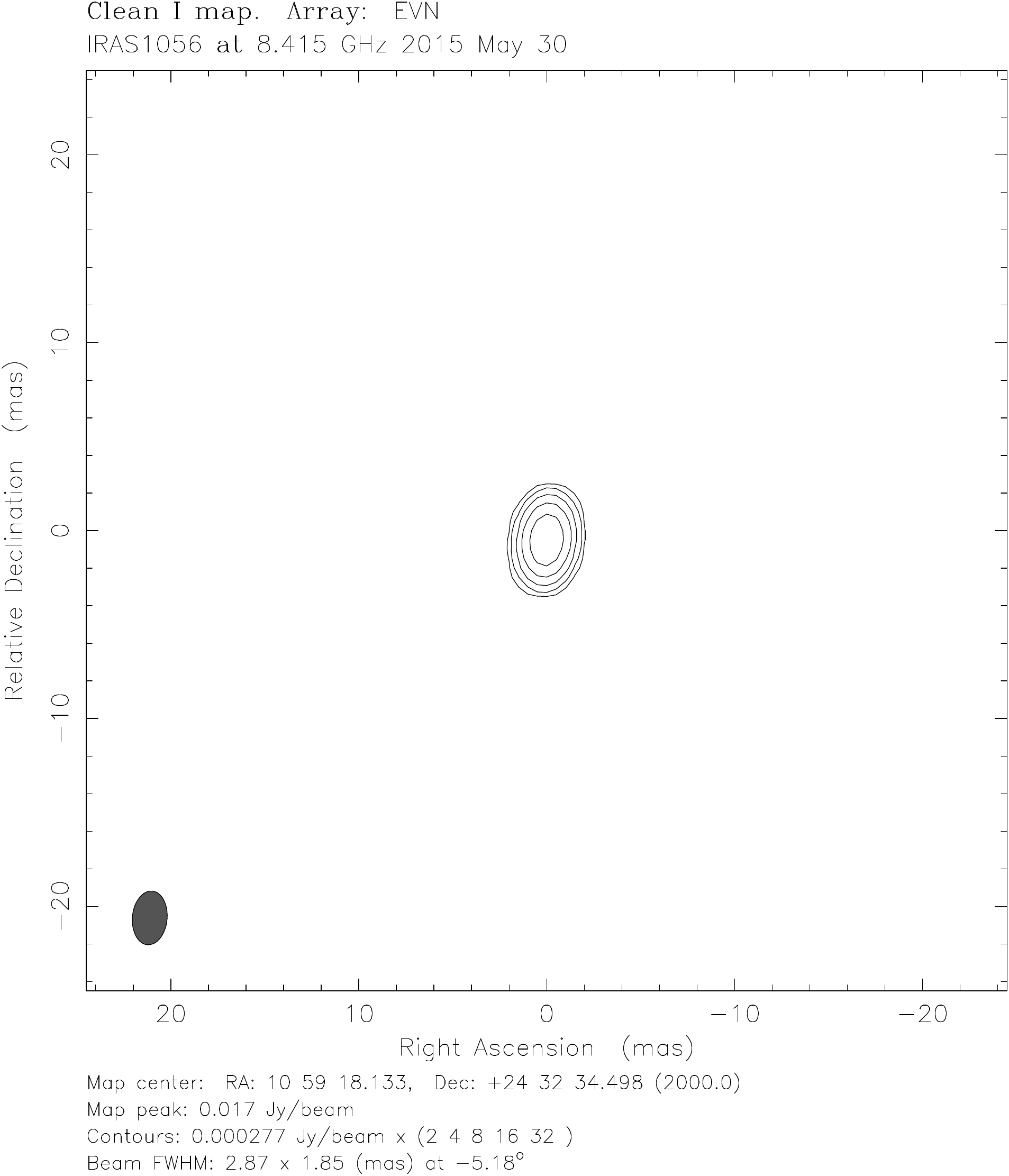} 
    \caption{The EVN VLBI images of IRAS 10565+2448 at 5 GHz (left) and 8.4 GHz (right) which have an rms noise $\sim 55.6\,\rm \mu Jy\,beam^{-1}$ and $\sim 142.7\,\rm \mu Jy\,beam^{-1}$ respectively. While there is only a core component in the 8.4 GHz image, the core-jet morphology elongated from northeast to southwest is clearly seen in the 5 GHz image.}
    \label{fig:evn_image}
\end{figure*}

Previous optical and X-ray observations found little evidence for the existence of an AGN in IRAS 10565+2448 \citep{veilleux1995,yuan2010,teng2010}. However, some radiatively inefficient AGN have little or no observational signature in the optical and X-ray bands \cite{heckman2014}. The presence of a jet in IRAS 10565+2448 is confirmed by observations from the European VLBI Network (EVN). We downloaded the pipeline-calibrated visibilities at 5 GHz and 8.4 GHz of IRAS 10565+2448 and produced the images in Difmap, which are showed in Fig. \ref{fig:evn_image}. We argue that the detected radio emission supports the existence of a jet based on the points below:

\begin{enumerate}
\item
The brightness temperature. Using the 8.4 GHz data and following the Equation 10 in \cite{shen1997}, we get a brightness temperature of $3.3\times10^{10}$ K that is much higher than those, $<10^{5}$ K, of starbursts \citep{condon1991}. Actually, radio emission from normal starbursts are generally resolved out in VLBI observations. 

\item
The morphology. In some extreme starburst galaxies, like Mrk 273 \citep{carilli2000,bondi2005} and Arp 220 \citep{smith1998,lonsdale2006,batejat2011,batejat2012,varenius2019}, there is indeed radio emission from starbursts detected by VLBI. However, the VLBI scale radio morphologies of those starbursts are very different from AGN jet morphologies. The radio morphologies of those extreme starbursts are a clumpy of compact radio sources, see the Figure 2 in \cite{carilli2000} and Figure 1 in \cite{bondi2005} for Mrk 273 and Figure 1 in \cite{smith1998}, Figure 1 in \cite{lonsdale2006}, Figure 1 in \cite{batejat2011}, and Figure 1 in \cite{batejat2012} for Arp 220. In IRAS 10565+2448, the EVN detected radio emission is however not one of clumpy compact radio sources.

\item
The radio luminosity.  From \cite{varenius2019} we know that in Arp 220 the average flux density of starburst radio emission at 5 GHz is 0.255 mJy, giving a luminosity of $1.9\times10^{20}\,\rm W\,Hz^{-1}$. From \cite{bondi2005}, we know that in Mrk 273 the brightest radio emission of starbursts has a peak flux density of 0.74 mJy\,beam$^{-1}$ at 5 GHz, giving a maximum luminosity of $2.4\times10^{21}\,\rm W\,Hz^{-1}$. In IRAS 10565+2448, EVN observations give a flux density of 24.6 mJy at 5 GHz, corresponding to a luminosity of $1.0\times10^{23}\,\rm W\,Hz^{-1}$ which is much higher than those from starbursts. 
\end{enumerate}
On this basis, we argue that the radio emission detected by the EVN observations comes from a jet.

\subsubsection{Energy- or momentum-conserving wind-driven or radiation-driven outflow?}

AGN-induced outflows are a major feedback mechanism in galaxies, and can be classified as energy-conserving mode, momentum-conserving mode, or radiation-pressure mode, depending on the physical environments surrounding the SMBH \citep[e.g.][]{silk1998,king2003,king2010,king2015,faucher2012,zubovas2012,fabian2012,costa2014}. 

In an energy-conserving outflow, the thermal energy of the AGN wind radiates away very slowly and so `keeps conserved' while the wind expels the surrounding interstellar medium (ISM). Assuming a SMBH accreting at the Eddington limit, a high gas covering fraction $\sim$1 and a 100 per cent thermal-to-kinetic conversion efficiency, the kinetic power of an energy-conserving outflow is about 5 per cent of the AGN bolometric luminosity \citep[e.g.][]{king2010,zubovas2012,costa2014}. 
In our target, the kinetic power of the outflows is at least $8.9\times10^{42}\,\rm erg\,s^{-1}$ and the AGN has a radiation power of $6.5\times10^{44}\,\rm erg\,s^{-1}$ if the AGN contributes 17 percent of the total bolometric luminosity of the galaxy \citep{veilleux2009,cicone2014}.  Thus the ratio between outflow kinetic power and AGN radiation power is $\ge0.014$. Although the lowest ratio is consistent with what expected from the `energy conserving' scenario, we argue that the outflow is not, or mainly, driven by the `energy conserving' wind based on the three points below:
\begin{enumerate}

\item 
As stated in the Section \ref{sec:multi_phase_outflows}, the mass outflow rate and outflow kinetic power we obtained are lower limits. If we assume the \mbox{H\,{\sc i}} gas spin temperature is 1000 K that is not high as there is already ionised outflow counterpart, then we would have total mass outflow rate of $\sim\,490\,\rm M_\odot \, yr^{-1}$ and outflow kinetic power of $\sim 3.2\times10^{43}\,\rm erg\,s^{-1}$. This kinetic power is just 4.9 percent of the AGN radiation power. It seems the AGN could drive the observed outflow. However, this is too optimistic because in reality the thermal-to-kinetic conversion efficiency and the gas covering fraction assumed above are upper limits and significant lower couplings are expected in some simulations \citep[e.g.][]{bourne2014,roos2015}, implying that much less than 5 percent of the AGN bolometric luminosity can be converted into the kinetic power of an outflow. Further, if we add the low velocity part of molecular outflow and the possible redshifted \mbox{H\,{\sc i}} outflow and  further correct the projection effect, we would obtain a higher outflow mass rate and kinetic power.

\item 
A more direct piece of evidence comes from the morphology of the ionised outflows. As stated in the Section \ref{sec:multi_phase_outflows}, the 1 kpc scale ionised outflow extending to southeast has higher FWHM indicated by $\rm H\alpha$ emission \citep{rupke2013}. The outflow driven by an `energy conserving' wind generally have a large opening angle. Thus if the 1 kpc scale ionised outflow is driven by an AGN wind, we should see the gas surrounding the nucleus have similar FWHM as the FWHM of the 1 kpc scale ionised outflow, which is not the case for the optical observations \citep{rupke2013}. Similarly, we only detected the \mbox{H\,{\sc i}} outflow in square 6, which also does not support the large-opening-angle outflow.

\item
Lastly, the velocity FWHM of the \mbox{H\,{\sc i}} outflow is large $\sim 581\,\rm km\,s^{-1}$ (noting that this is just the projected quantity to line of sight), and is very similar to those produced by jet-gas interactions as noted in previous studies \citep[e.g.][]{morganti2005,mahony2013}.
\end{enumerate}

In contrast to an `energy conserving' outflow, if the thermal energy of the AGN wind radiates away very quickly during its expansion, then the energy is not conserved but the momentum is still conserved. This corresponds to a `momentum-conserved' outflow, which is less efficient than the `energy conserving' outflow and the outflow kinetic power is expected to be about 0.1 per cent of the AGN radiation power \citep[e.g.][]{king2015}. Thus the outflows in IRAS 10565+2448 are not driven by AGN momentum. One property of an AGN momentum-driven outflow is that the outflow momentum rate is close to the AGN radiation momentum rate. In IRAS 10565+2488, the momentum rate of the outflows is at least $\dot{M}V =3.3\times10^{35}\,\rm g\,cm\,s^{-2}$ which is around 14.1 times the AGN momentum rate $L_{\rm AGN}/c$, where $c$ is the speed of light. 

The AGN radiation can directly interact with dust clouds in the ISM and thus drive outflows \citep[e.g.][]{fabian2012,ishibashi2015,ishibashi2017,ishibashi2018}. In this scenario, the outflow kinetic power is expected to be about 1 per cent of the AGN radiation power and the outflow momentum rate is expected to be up to 5 times $L_{\rm AGN}/c$. Based on the observed properties of the outflows, the radiation-pressure mode does not appear to explain them.

Therefore, while we could not exclude the existence of AGN-wind (radiation)-driven outflows, the AGN wind (radiation) should make only a minor contribution to driving the outflows.

\subsubsection{Jet-driven outflow?}\label{sec:jet_driven_outflow}
The final mechanism that can drive massive gas outflows is the radio jet. During the growth of a radio jet it can clear and press the gas on its propagation way and thus drive outflows, and this model is supported by the high spatial resolution observational results that the outflows are found to be at the hot spots of the jet \citep[e.g.][]{morganti2013,mahony2013,murthy2022}. From hydrodynamical simulations, a radio jet is capable of accelerating clouds to high velocities if the ratio of jet power to Eddington luminosity $\eta = P_{\rm jet}/L_{\rm Edd}$ is above $10^{-4}$ \citep[e.g.][]{wagner2011,wagner2012,mukherjee2018}. 

For our target, the integral flux density observed in our uGMRT image is 49.6\,mJy at 1.36 GHz which corresponds to a  radio luminosity of $\sim 2.1\times10^{23}\,{\rm W\,Hz^{-1}}$ at 1.40 GHz in the rest frame. However, the lower-resolution NRAO VLA Sky Survey (NVSS) gives a total 1.4 GHz flux density of 57.6 mJy \citep{condon1998}, implying a slightly higher 1.4 GHz radio luminosity of $2.5\times10^{23}\,{\rm W\,Hz^{-1}}$. We note some radio emission may come from star formation. If we use the star formation rate in our target and the equation 3 in \cite{sullivan2001}, we estimated the maximum 1.4 GHz radio luminosity contribution from star formation is $1.2\times10^{23}\,{\rm W\,Hz^{-1}}$. Thus, we still have $1.3\times10^{23}\,{\rm W\,Hz^{-1}}$ from a jet, corresponding to a jet power $5.3\times10^{42}\,\rm erg\,s^{-1}$ using the Equation 2 in  \mbox{\cite{best2006}}. The estimated SMBH mass of $\sim 2.0\times10^{7}\,{\rm M_\odot}$ implies an Eddington luminosity of $\sim 2.6\times10^{45}\,\rm erg\,s^{-1}$ \mbox{\citep{dasyra2006}}. Then the ratio of jet power to Eddington luminosity is $\sim2.0\times10^{-3}$ which is larger than $10^{-4}$.

Therefore, the radio jet in IRAS 10565+2448 can drive the observed outflows. More evidence in support of a jet-driven outflow comes from the morphology of the radio emission and the spatial distribution of the \mbox{H\,{\sc i}} and molecular outflows. A Very Large Array (VLA) continuum image at 8.4 GHz with a resolution of 0.25 arcsec \citep{condon1991} shows that the radio emission is elongated from northeast to southwest. This spatial alignment between the \mbox{H\,{\sc i}} and molecular outflows and the radio emission supports the jet-driven scenario. Additionally, the velocity width of the outflows is very broad which is a common feature of jet-gas interactions such as that observed in e.g. 3C 293 \citep{mahony2016}, 4C 12.50 \citep{morganti2013}, and 3C 236 \citep{schulz2018}.

Based on the discussion above, we argue that the radio jet is capable of driving the observed outflows in addition to the outflows driven by the starburst.

\subsection{Implications}
It is generally accepted that outflows, especially AGN-driven outflows, play a significant role in galaxy evolution. Quantifying the basic parameters regarding mass outflow rate, outflow mass, kinetic power, radius, spatial distribution, and momentum rate of outflows are crucial for understanding their impact and driving mechanism. However, this is often difficult as outflows contain multi-phase gas which requires multi-band observations. In IRAS 10565+2448, the optical warm ionised outflow has a mass outflow rate of $\sim 1.4\,{\rm M_\odot \, yr^{-1}}$ which is much smaller than that of the cold molecular outflow and neutral hydrogen outflow. Without the latter two detections, the impact of the outflow in IRAS 10565+2448 would be highly underestimated and the outflow driving mechanism would remain highly uncertain. Previous observations have also shown that the dominant mass in outflows is often in neutral or molecular \citep[e.g.][]{tadhunter2014,morganti2018}, which highlights the necessity of multi-band observations of outflows.      

High-resolution observations can help us directly image the outflows and thus determine the outflows radius. Given the outflow mass, if the outflowing gas is uniformly distributed within the outflow cone, the mass outflow rate is given by $\dot{M} = 3MV/R $, where $M$ is the outflow mass, $V$ the outflow velocity, and $R$ the outflow radius. Instead, if the outflowing gas is distributed in a thin shocked shell, the mass outflow rate would be reduced by a factor of three. 
In our target, we did not find a broad absorption wing towards nucleus or at least the broad absorption strength from square 5 is not as strong as that from square 6, which supports the shell-like outflow scenario. Another key element in outflows is the radius, the distance between nuclei and outflows. In previous low spatial resolution observations from which no accurate radius can be derived, people have to assume a value, e.g. 1 kpc. However, we can directly measure it and thus accurately measure the mass outflow rate if the outflows are spatially resolved. Therefore, high spatial resolution observations are crucial in quantifying the impact of outflows.     

Jets play a significant role in AGN feedback. While there is no doubt that bright jets can drive massive outflows and have powerful impact on host galaxies \citep[e.g.][]{morganti2018}, the influence of less powerful (`modest') jets on host galaxies has not been well recognised. A recent notable study of B2 0258+35 showed that the modest jet of $L_{\rm1.4GHz}\sim2.1\times10^{23}\,{\rm W\,Hz^{-1}} $ has carried $\sim 75\%$ gas in its central region \citep{murthy2022}, showing that the contribution from jets cannot be ignored even in the case of a modest jet like the one in B2 0258+35. Our target also has a jet with 1.4\,GHz radio luminosity of $\sim1.3\times10^{23}\,{\rm W\,Hz^{-1}}$ which reinforces this point. To make it clear whether jets contribute substantially to AGN-driven outflows, and whether jet-driven outflows are more powerful than AGN-wind (radiation)-driven outflows, we suggest that a larger AGN sample with clear jet evidence and a comparison AGN sample without jets are needed to enable a statistical study. 

\section{Summary}
We have presented uGMRT \mbox{H\,{\sc i}} spectral-line observations of the ULIRG IRAS 10565+2448, which was reported to show blueshifted, broad, and shallow \mbox{H\,{\sc i}} absorption indicating outflow and faint \mbox{H\,{\sc i}} emission in previous Arecibo observations \citep{mirabel1988}. The much higher spatial resolution of the uGMRT allowed us to separate the \mbox{H\,{\sc i}} absorption from emission and to localise the \mbox{H\,{\sc i}} outflow gas.

We did not detect the \mbox{H\,{\sc i}} emission see by Arecibo, since our observations have lower sensitivity and also much higher spatial resolution, which is likely to have resolved out  the faint \mbox{H\,{\sc i}} emission. 

We successfully recovered the blueshifted, broad, and shallow \mbox{H\,{\sc i}} absorption with rest-frame velocity up to $\sim -530$ km\,s$^{-1}$ which is slightly different from the maximum shifted velocity, $\sim -600$ km\,s$^{-1}$, in the Arecibo spectrum. This difference may result from our lower sensitivity. 

Besides the blueshifted \mbox{H\,{\sc i}} outflow, we made a possible detection of a redshifted \mbox{H\,{\sc i}} outflow in the opposite direction. By combining our \mbox{H\,{\sc i}} data with data on the warm ionised and cold molecular outflows \citep[e.g.][]{rupke2013,cicone2014}, the outflows in IRAS 10565+2448 have a total mass outflow rate of at least 140 $ \rm M_\odot \,yr^{-1}$ and a total energy loss rate of at least $8.9\times10^{42}\,\rm erg\,s^{-1}$. By analysing these data, we conclude that the radio jet may play a role in driving the observed outflows.

We stress the importance of multi-band observations to probe multi-phase gas and the necessity of high spatial resolution observations to image outflows, which can help us to accurately measure the properties of outflows and thus understand their driving mechanism and impact on the host galaxies. The modest jet radio luminosity, $L_{\rm1.4GHz}$ $\sim1.3\times10^{23}\,{\rm W\,Hz^{-1}}$, of IRAS 10565+2448 emphasises that we cannot ignore the contribution from radio jets in driving outflows even when the radio luminosity is modest, see also the case of B2 0258+35 \citep{murthy2022}. Finally, as one of few sources with spatially resolved optical, atomic, and molecular outflow detections, IRAS 10565+2448 can provide new constraints on understanding jet-gas interaction and inform future simulations of feedback processed in galaxy evolution.      
\section*{Acknowledgements}

We thank the anonymous referee for the comments that improve this paper.

We thank the staff of the GMRT who have made these observations possible. The GMRT is run by the National Centre for Radio Astrophysics of the Tata Institute of Fundamental Research.

RZS thanks CSIRO for their hospitality during his stay in Australia and acknowledges support from a joint SKA PhD scholarship.

RZS and MFG are supported by the National Science Foundation of China (grant 11873073), Shanghai Pilot Program for Basic Research--Chinese Academy of Science, Shanghai Branch (JCYJ-SHFY-2021-013), and the science research grants from the China Manned Space Project with NO. CMSCSST-2021-A06. This work is supported by the Original Innovation Program of the Chinese Academy of Sciences (E085021002).

Y.Chandola acknowledges the support from the NSFC under grant No. 12050410259, and Center for Astronomical Mega-Science, Chinese Academy of Sciences, for the FAST distinguished young researcher fellowship (19-FAST-02), and MOST for the grant no. QNJ2021061003L.

Parts of this research were supported by the Australian Research Council Centre of Excellence for All Sky Astrophysics in 3 Dimensions (ASTRO 3D), through project number CE170100013. 
MG (Curtin) was partially supported by the Australian Government through the Australian Research Council's Discovery Projects funding scheme (DP210102103).

\section{DATA AVAILABILITY}
The EVN data are available through its archive. Other data are available on reasonable request to the corresponding author.

\bibliographystyle{mnras}
\bibliography{iras10565} 





\bsp	
\label{lastpage}
\end{document}